%% file: 38019corr_max.tex
\documentclass[traditabstract]{aa} %
\usepackage{graphicx}
 \usepackage{epstopdf}
\usepackage{color}
\usepackage{savesym}
\usepackage{deluxetable} 
\restoresymbol{DTBL}{rotate}
\usepackage{txfonts} 
\usepackage{natbib}
\usepackage{graphicx}
\bibpunct{(}{)}{;}{a}{}{,} 
\usepackage[english]{babel}
\usepackage{url}
\usepackage{hyperref}

\begin{document} 

\title{The Carnegie Supernova Project II. Observations of the luminous red nova  AT~2014ej\thanks{This paper includes data gathered with the 6.5 meter Magellan telescopes at Las Campanas Observatory, Chile.}$^{,}$\thanks{Photometry and spectra presented in this paper are available on \href{https://wiserep.weizmann.ac.il/}{WISeREP}.}}

\author{M. D. Stritzinger\inst{1}
\and F. Taddia\inst{1}
\and M. Fraser\inst{2}
\and T.~M. Tauris\inst{3,1}
\and C. Contreras\inst{1,4}
\and S Drybye\inst{1,5}
\and L. Galbany\inst{6}
\and S. Holmbo\inst{1}
\and N. Morrell\inst{4}
\and A. Pastorello\inst{7}
\and M. M. Phillips\inst{4}
\and G. Pignata\inst{8,9}
\and L. Tartaglia\inst{10}
\and N. B. Suntzeff\inst{11,12}
\and J. Anais\inst{4}
\and C. Ashall\inst{13}
\and E. Baron\inst{14}
\and C. R. Burns\inst{15}
\and P. Hoeflich\inst{13}
\and E. Y. Hsiao\inst{13}
\and E. Karamehmetoglu\inst{1}
\and T. J. Moriya\inst{16,17}
\and G. Bock\inst{18}
\and A. Campillay\inst{4}
\and S. Castell\'{o}n\inst{4}
\and C. Inserra\inst{19}
\and C. Gonz\'{a}lez\inst{4}
\and P. Marples\inst{20}
\and S. Parker\inst{21}
\and D. Reichart\inst{22}
\and S. Torres-Robledo\inst{23,4}
\and  D. R. Young\inst{24}
}

\institute{Department of Physics and Astronomy, Aarhus University, 
Ny Munkegade 120, DK-8000 Aarhus C, Denmark 
\and 
School of Physics, O'Brien Centre for Science North, University College Dublin, Belfield, Dublin 4, Ireland
\and
Aarhus Institute of Advanced Studies (AIAS), Aarhus University, H{\o}egh-Guldbergs Gade 6B, 8000 Aarhus C, Denmark
\and
Las Campanas Observatory, Carnegie Observatories, Casilla 601, La Serena, Chile 
\and
Nordic Optical Telescope, Apartado 474, E-38700 Santa Cruz de La Palma, Spain
\and
Departamento de F\'isica Te\'orica y del Cosmos, Universidad de Granada, E-18071 Granada, Spain
\and
INAF - Osservatorio Astronomico di Padova, Vicolo dell'Osservatorio 5, 35122, Padova, Italy
\and
Departamento de Ciencias Fisicas, Universidad Andres Bello, Avda. Republica 252, Santiago, Chile
\and
Millennium Institute of Astrophysics, Santiago, Chile
\and
The Oskar Klein Centre, Department of Astronomy, Stockholm University, AlbaNova, 10691 Stockholm, Sweden
\and
The George P. and Cynthia Woods Mitchell Institute for Fundamental Physics and Astronomy, Texas A\&M University, College Station, TX 877843
\and
Department of Physics and Astronomy, Texas A\&M University, College Station, TX 77843
\and
Department of Physics, Florida State University, Tallahassee, FL 32306, USA
\and
Homer L. Dodge Department of Physics and Astronomy, University of Oklahoma, 440 W. Brooks, Rm 100, Norman, OK 73019-2061, USA
\and
Observatories of the Carnegie Institution for Science, 813 Santa Barbara St, Pasadena, CA, 91101, USA
\and
National Astronomical Observatory of Japan, National Institutes of Natural Sciences, 2-21-1 Osawa, Mitaka, Tokyo 181-8588, Japan
\and
School of Physics and Astronomy, Faculty of Science, Monash University, Clayton, VIC 3800, Australia
\and
Runaway Bay, Gold Coast, Queensland, Australia
\and
School of Physics \& Astronomy, Cardiff University, Queens Buildings, The Parade, Cardiff, CF24 3AA, UK
\and
Leyburn Observatory, Queensland 4129, Australia.
\and
Parkdales Observatory, Oxford, New Zealand
\and
University of North Carolina at Chapel Hill, Campus Box 3255, Chapel Hill, NC 27599-3255, USA 
\and
SOAR Telescope, La Serena 1700000, Chile
\and
Astrophysics Research Centre, School of Mathematics and Physics, Queen's University Belfast, Belfast BT7 1NN, UK} 

  \abstract
   {We present  optical and near-infrared  broadband  photometry and optical spectra of AT~2014ej from the  Carnegie Supernova Project-II. These observations are  complemented with  data  from  the CHilean Automatic Supernova sEarch, the Public ESO Spectroscopic Survey of Transient Objects, and from  the Backyard Observatory Supernova Search. 
   Observational signatures of AT~2014ej  reveal that it is similar to other members of the gap-transient subclass known as luminous red novae (LRNe), including the ubiquitous double-hump light curve and  spectral properties similar to that of LRN SN~2017jfs. 
   A medium-dispersion visual-wavelength spectrum  of AT~2014ej taken with the Magellan Clay telescope exhibits a P~Cygni H$\alpha$ feature characterized by a blue velocity at zero intensity of $\approx 110$~km~s$^{-1}$ and a P~Cygni minimum velocity of $\approx  70$~km~s$^{-1}$. We attribute this to emission from a  circumstellar wind. 
   Inspection of pre-outbust  \textit{Hubble} Space Telescope images   yields no conclusive progenitor detection.   
  In comparison with a sample of LRNe from the literature, AT~2014ej lies at the brighter end of the luminosity distribution.
  Comparison of the ultra-iolet, optical, infrared (UVOIR)  light curves of well-observed LRNe   to   common-envelope evolution models from the literature indicates that the models underpredict the luminosity of the comparison sample at all phases and also produce inconsistent timescales of the secondary peak.    Future  efforts to model LRNe should expand upon the  current parameter space we explore here and  therefore may consider more massive systems and a wider range of dynamical timescales.}
   \keywords{star: mass loss, circumstellar matter - supernova: individual AT~2014ej  (PSN J23160979$-$4234575)}
\authorrunning{Stritzinger, Taddia, Fraser, et al.}     
\titlerunning{CSP-II observations of the LRN AT~2014ej}    

 \maketitle
 
\section{Introduction}

  This is the second of two papers by the \textit{Carnegie Supernova Project-II} (CSP-II) that each present a case study of a gap transient.  Gap transients are located within the  so-called ``luminosity gap''  parameter space  that is devoid of bright novae ($M_V \lesssim -10$ mag) and the least luminous core-collapse supernovae (SN; peak $M_V \sim -16$ mag).
  Because the visual wavelength spectra of  gap transients (in the past also often referred to as  intermediate luminosity optical transients) typically resemble the spectra of interacting type~IIn SNe, they were historically commonly called SN imposters \citep[e.g.,][]{1990MNRAS.244..269S,1997ARA&A..35..309F,2000PASP..112.1532V}. 
  As the population of gap transients with decent observational data  has increased, it has become apparent that the  diversity displayed among the literature sample suggests distinctly different populations \citep[see][for more detailed discussion]{2009aaxo.conf..312K,2012ApJ...758..142K,kashi2016,2019NatAs...3..676P}. 
  
 Today,  at least three known populations of gap transients are recognized in the luminoisty gap. These include classical luminous blue variable (LBV) outbursts, intermediate-luminosity red transients (ILRTs), and luminous red novae (LRNe). LBVs are thought to be related to eruptions of massive luminous stars \citep[see][]{smith11}. As shown in greater detail in \citet[][hereafter  Paper~1]{Stritzinger2020}, the ILRT subtype is well represented by NGC~300-2008-OT and SN~2008S, and has been linked to  asymptotic giant branch (S-AGB) stars \citep{2008ApJ...681L...9P,thompson09,2009MNRAS.398.1041B,prieto2009,2011ApJ...741...37K,2016MNRAS.460.1645A,2017PASA...34...56D} that die as electron-capture supernovae  \citep{1980PASJ...32..303M,1984ApJ...277..791N,1987ApJ...318..307M,1993ApJ...414L.105H,2006A&A...450..345K,2008ApJ...675..614P}.  Other models  appearing in the literature for ILRTs consist of moderately massive stars experiencing super-Eddington winds and/or giant outbursts \citep[e.g.,][]{2009ApJ...697L..49S,2011ApJ...743..118H}, or massive stars donating material to a main-sequence star, leading to the release of gravitational energy  \citep[e.g.,][]{kashi2010}.   Finally, a leading model for the origins of LRNe,  which all display a ubiquitous double-humped light curve  \citep{2019A&A...630A..75P}, consists of the ejection of a common envelope by a massive binary system  \citep[e.g.,][]{2017ApJ...834..107B} upon coalescence \citep{2016MNRAS.458..950S,2017MNRAS.471.3200M,2017MNRAS.470.2339L,2018MNRAS.473.3765M}. However, other models have also been proposed in the past to account for LRNe, particularly within articles that have studied the Galactic LRN archetype V838~Mon. These include, among others, outbursts from massive stars \citep{tylenda05},  accretion of low-mass stars onto solar-mass main-sequence companions \citep{soker2003,tylenda2006,kashi2010,kashi2016,soker2020}, or even giant stars that   accrete  relatively massive  planets \citep{retter2003}.  
  
The locations of the three gap-transient subtypes in the luminosity versus decay time (defined as the time interval in which the $r$-band light curve drops  one magnitude from peak) parameter space are shown in Fig.~1 of  Paper~1. 
  In general, the various gap-transient subtypes  exhibit similar peak luminosities, while differences among their light curve decay times are apparent. LBVs tend to exhibit a more slowly  declining light-curve evolution. For example, as indicated in Fig.~1 of Paper~1,  LBVs typically exhibit decay timescales   of $\gtrsim$10${^2}$~days. On the other hand, both ILRTs and LRNe typically decay on shorter  timescales of 10$^{1}-10^{2}$~days.   Because very few gap transients have been studied in depth, the full extent of  the peak luminosity versus decay time parameter space populated by both ILRTs and LRNe is currently unknown.
  
Similar to the ILRT SNhunt120 presented in Paper~1, here in Paper 2, we add to the small but growing sample of LRNe with our observational data set of AT~2014ej. The data set is based largely on observations performed by the CSP-II \citep{phillips2019}. This includes optical light curves and some near-IR (NIR) photometry, as well as one  low- and  one medium-dispersion visual wavelength spectrum.  The CSP-II data are clearly complemented with unfiltered images from the Backyard Observatory Supernova Search\footnote{\url{https://www.bosssupernova.com/}} (BOSS) and the CHilean Automatic Supernova sEarch (CHASE; \citealt{2005NCimC..28..767R}), as well as a handful of visual wavelength spectra and optical/NIR broadband photometric measurements obtained by   the Public ESO Spectroscopic Survey of Transient Objects (PESSTO; \citealt{smartt2015}).
In addition, archival images of the host galaxy NGC~7552  obtained with the \textit{Hubble} Space Telescope (HST) are examined in order to compute progenitor limits. 

\section{AT~2014ej} 
\subsection{Discovery, distance, and reddening} 
\label{AT2014ejdetails}

PSN J23160979-4234575 (hereafter we refer to this transient by its Transient Name Server designation AT~2014ej) was discovered in NGC~7552 on 2014 September 24.46 UT by Peter Marples during the course of the BOSS  \citep{2014CBET.3998....1B} and was confirmed to be $\sim$ 17.8$\pm$0.4 magnitude in unfiltered image taken by G. Bock on 2014 September 24.5 UT.
\citeauthor{2014CBET.3998....1B} also reported nondetections from images with a limiting magnitude of $\sim 18.5$ taken on 2.46 and 2014 September 10.50 2014 UT.
We reexamined the discovery image, and after removing the host at the position of the transient, we computed photometry relative to $r$-band photometry of a local sequence of stars in the field of NGC~7552 (see Sect.~\ref{AT2014ejobservations}). This provides an apparent magnitude  of  $17.96\pm0.11$ mag on 2014 September 24 (MJD 56923.66), which is slightly fainter than the value reported \citep{2014CBET.3998....1B}. Furthermore, we compute a 3-sigma nondetection limit of 17.8 mag for an image obtained by BOSS on the previous night.

\begin{figure}
\centering
\includegraphics[width=12cm]{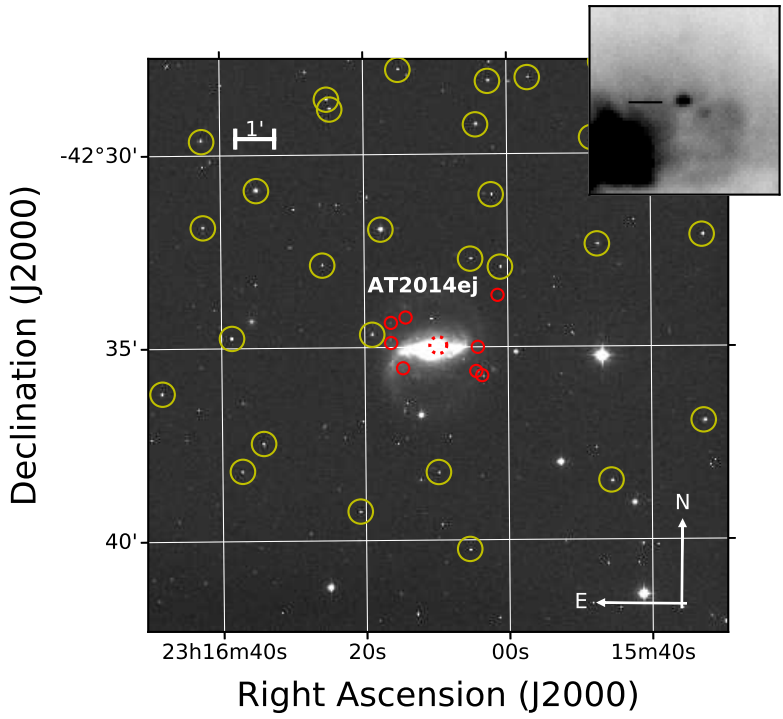}
\caption{ Finding chart of  NGC~7552 hosting AT~2014ej  constructed from a single Swope $r$-band image.
The position of the transient  is indicated with a dotted circle with a zoom-in of the area enclosed shown in the upper right corner inset. Optical and NIR local sequence stars are indicated with yellow and red circles, respectively.} 
\label{fig:FC2}
\end{figure}

As  shown below,  AT~2014ej was recovered in unfiltered search images taken by CHASE about four days prior to the BOSS detection on 2014 September 20.06 with an apparent magnitude of $17.5\pm0.1$.
The J2000 coordinates of AT~2014ej are $\alpha=23^{\rm h}16^{\rm m}09\fs79$ and $\delta=-42^{\circ}34\arcmin57\fs5$, which is
$10\farcs9$ west and $7\farcs6$ north  from the core of its SBab host galaxy NGC~7552. A finding chart of NGC~7552 is shown in  Fig.~\ref{fig:FC2}.

\citet{2014ATel.6508....1M} obtained a visual wavelength spectrum  with the du Pont telescope located at the Las Campanas Observatory (LCO) on 2014 September 27.2 UT and classified AT~2014ej as a SN imposter. This early spectrum exhibits Balmer line velocities of $\approx$~800 km~s$^{-1}$. As the transient appeared to be caught on the rise, the CSP-II decided to include it within its followup program. 

According to NED, NGC~7552 is located at a heliocentric redshift of $z = 0.00537\pm0.00002$ (1608 km~s$^{-1}$). 
To convert this redshift into distance, we adopted the following cosmological parameters: $H_0 = 73.2\pm2.3$ km~s$^{-1}$~Mpc$^{-1}$ \citep{2018ApJ...869...56B}, $\Omega_m = 0.27$, and $\Omega_L = 0.73$, as well as a correction from NED based on Virgo, Great Attractor, and Shapley in-flow models. According to this, the redshift distance corresponds to a distance modulus of  $\mu = 31.72\pm0.15$ mag and hence the distance  $D$~$=$~$22.07\pm1.53$ Mpc. 

As NGC~7552 is a member of the Grus Quartet, we turn to distance estimates of the group itself as well as to the individual  members NGC 7582, NGC~7590, and NGC~7599.
NED lists five different Tully-Fisher distances for NGC~7582, with the most recent being $\mu = 31.76\pm0.20$ mag. NED also lists a handful of Tully-Fisher distances to NGC~5790, with the most recent being from the Cosmicflows-2 catalog \citep{tully2013} , corresponding to  $\mu = 32.46\pm0.20$ mag. Finally, NED also lists a variety of Tully-Fisher distances for NGC~7599, with the two most recent values coming in as $\mu = 31.22\pm0.20$ mag from the Cosmicflows-2 catalog \citep{tully2013} and $\mu =  31.02\pm0.45$ mag from the Cosmicflows-3 catalog  \citep{tully2016}.

NGC~7582 and NGC~7552 are interacting, as is shown by tidal extension of HI between the galaxies \citep{grus_quartet}.
The  distance modulus of NGC~7582 is  $\mu = 31.76\pm0.20$ mag,  in agreement with the distance modulus of NGC~7552 as computed from its redshift.
 In the following, we adopt $\mu = 31.72\pm0.15$ mag as the distance modulus to NGC~7552.
 
Finally, in Appendix~\ref{appendixA} we compute the host  metallicity using a spectrum of NGC~5775 obtained by the Sloan Digital Sky Survey (SDSS), while Appendix~\ref{appendixB} contains details on the total reddening value used in this work.  In short, we find the metallicity at the location of AT~2014ej to be super-solar, and our best estimate for  the total visual extinction of AT~2014ej is  $A^{tot}_{V} = 0.96\pm0.47$ mag.

\subsection{Observations}
\label{AT2014ejobservations}

The CSP-II obtained 20 epochs of optical $ugriBV$-band imaging with the Swope (+ CCD camera) telescope, extending between $+$5~d to $+$37~d relative to first detection.\footnote{First detection of AT~2014j was made by CHASE and occurred on JD-2456920.56.} In addition, three epochs of NIR $YJH$-band imaging were obtained with the du Pont (+ RetroCam) telescope between $+$9~d to $+$12~d, as well as a single epoch of $K_s$-band photometry taken with the Magellan Baade telescope equipped with the FourStar imager \citep{2013PASP..125..654P} on $+$17~d. 

CSP-II photometry of the transient was computed relative to an optical/NIR local sequence consisting of 27/7 stars, themselves  calibrated relative to standard star fields observed over multiple photometric nights. Optical ($ugriBV$) and NIR ($YJH$) photometry of the local sequence stars in the `standard' system are listed in Table~\ref{tab:CSPoptlocseq} and  Table~\ref{tab:AT2014ejnirlocseq}, respectively. 
Optical photometry of AT~2014ej in the `natural' photometric system is listed in  Table~\ref{CSPII-optphot}, while Table~\ref{tab:AT2014ej_nirphot} lists the NIR photometry also in the CSP-II natural system.
 Finally, a single $K_s$-band epoch of photometry was calibrated relative to 2MASS stars in the field of NGC~7552 and is also listed in Table~\ref{tab:AT2014ej_nirphot}. 

AT~2014ej was observed by CHASE \citep{2009RMxAC..35R.317P} with the PROMPT telescopes \citep{2005NCimC..28..767R} located at CTIO. We detect AT~2014ej in 14 epochs of unfiltered images. This includes the earliest detection and the  subsequent detection in survey search images extending over  $+$95~d.
The images were reduced following standard procedures, and photometry of AT~2014ej was computed relative to the $r$-band photometry of the CSP-II local sequence of stars in the field of the host galaxy. 
Prior to computing photometry, a deep host-galaxy template image was constructed and used to subtract the host light at the position of the transient. The stacked images were obtained prior to the discovery of AT~2014ej. 
Unfiltered photometry obtained with the PROMPT telescopes is listed in Table~\ref{tab:AT2014ej_CHASE_phot}.

PESSTO obtained a handful of $JHK$-band images with the NIR imager SOFI (Son OF ISAAC; \citealt{moorwood1998}) attached to the ESO New Technology Telescope (NTT)\footnote{See details in the PESSTO third data release documentation located at the url \url{http://www.eso.org/rm/api/v1/public/releaseDescriptions/88}.}.  Point-spread function (PSF) photometry of the transient was calibrated relative to 2MASS stars in the field of NGC~7552, and the results are listed in Table~\ref{tab:AT2014ej_nirphot_pessto}. This includes a single epoch of $JH$-band photometry and three epochs of $K$-band photometry.

Five epochs of low-resolution visual wavelength spectroscopy of AT~2014ej were obtained between $+$7~d to $+$72~d, and these observations are summarized in Table~\ref{tab:specobs}. These spectra include data taken by the CSP-II with the du Pont ($+$ WFCCD) telescope and by PESSTO with the NTT (+ EFOSC2) and a previously published spectrum obtained with the ANU telescope ($+$ WiFes; \citealt{childress2016}). The spectroscopic data were reduced following standard procedures.  In addition to these low-resolution spectra, a single medium-dispersion spectrum was obtained on $+$11~d with the 6.5m Magellan Clay telescope equipped with MIKE (Magellan Inamori Kyocera Echelle; \citealt{2003SPIE.4841.1694B}). The MIKE spectrum followed standard procedures and made use  of \texttt{IRAF}\footnote{IRAF is distributed by the National Optical Astronomy Observatory, which is operated by the Association of Universities for Research in Astronomy (AURA) under cooperative agreement with the National Science Foundation.} echelle routines and the \texttt{mtools} standard package developed by  Jack Baldwin.

Finally, we note that the $V$-band acquisition images taken in the process of obtaining  spectroscopic followup of  AT~2014ej with the NTT were used to measure $V$-band photometry in four different epochs, three of which occurred after the end of the CSP-II NIR followup. PSF photometry of AT~2014ej was computed from these images and calibrated relative to the CSP-II $V$-band local sequence. The resulting photometry is listed in Table~\ref{tab:AT2014ej_pessto_phot}. 
With the $V$ band from CSP-II and PESSTO acquisition images, we performed an absolute flux calibration of  the spectra, and from these, we obtained spectrophotometry in the $Bgri$ bands. Because the host galaxy flux might be contaminated, we assumed
a conservative 0.1~mag error on spectrophotometry, which is listed in Table~\ref{tab:spectrophotometry}. Spectrophotometry was useful to cover $Bgi$ bands at epochs after $+$24~d.
 
\section{Results}
\subsection{Photometry, broadband colors, and SEDs}
\label{AT2014ejphotometry}

\begin{figure}
   \includegraphics[width=18cm]{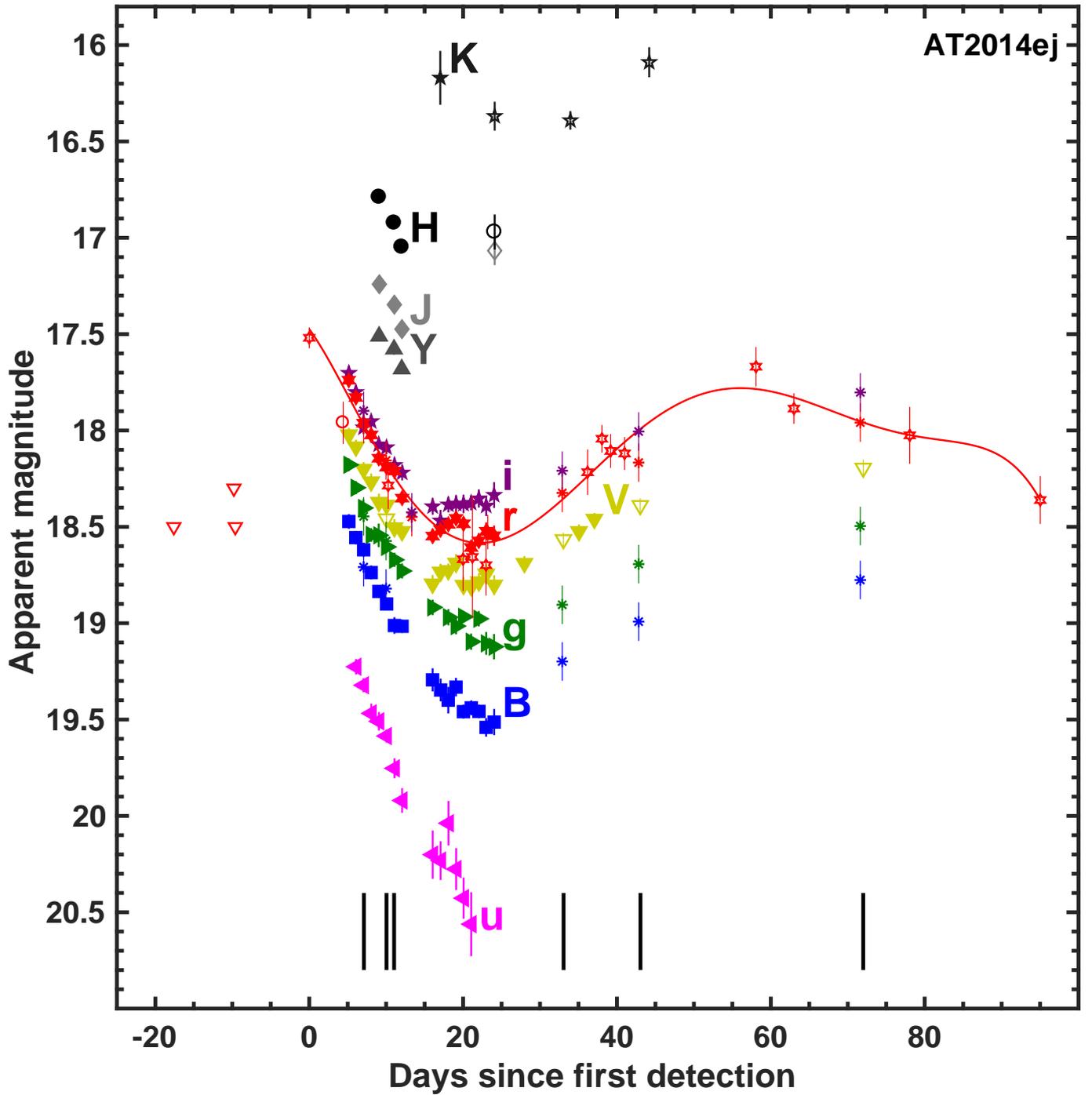}
   \caption{Optical and NIR light curves of AT~2014ej.  The light curves decline rapidly over the first $\sim$20 days when they appear to reach a minimum and then evolve to a second peak (around 60 days in $r$ band). The open red circle is the BOSS discovery magnitude, while open red triangles at 18.5 mag correspond to nondetection limits reported by \citet{2014CBET.3998....1B}, and the open red triangle at 18.3 is a nondetection limit determined from an unfiltered CHASE image. Open red stars are CHASE photometry, and open yellow triangles and open symbols in the NIR bands are PESSTO (EFOSC2 and SOFI) photometry ($V$, $J$, $H$, $K$). Colored asterisks are spectrophometry ($B$, $g$, $r$, $i$). The other colored filled markers are CSP-II photometry. The optical $r$-band light curve is fit with a low-order polynomial shown as a solid red line. Optical spectral phases are marked by black segments.}
\label{Fig:AT2014ejphotometry}
    \end{figure}

Photometry of AT~2014ej is plotted in Fig.~\ref{Fig:AT2014ejphotometry}.
Post detection, the optical light curves decline in brightness, exhibiting a drop between 0.5 to 1.0 mag over the first ten days.  This decline continues in all bands out to $\geq$ $+$15~d. 
Subsequently, as the $u$ band continues to exhibit a rapid drop in brightness, the $B$- and $g$-band light curves significantly slow down in their rate of decline, while the $V$-, $r$-, and $i$-band light curves exhibit a plateau phase that last  about ten days. In the CHASE $r$-band data, the PESSTO  $V$-band images, and spectrophotometry in $Bgri$ bands, we  observe a rise in the light curves from each of these optical bands after +24d. The rise in the $r$ band goes from 18.5 mag up to 17.7 mag between $+$24~d and $+$60~d, followed by a shallow decline reaching 18.4 mag at $+$95~d.

The first CHASE detection of AT~2014ej reveals a peak apparent magnitude of $\leq$17.5, corresponding to a peak absolute magnitude of $M_r \leq-15.09\pm0.48$, given the assumed extinction and distance modulus.
The accompanying uncertainty is computed by adding in quadrature the errors in adopted reddening and distance. 
Taking an average between the dates of last nondetection and first detection, we estimate an outburst date of  JD=2456915.75$\pm$4.80.

\begin{figure}
\centering
\includegraphics[width=10cm]{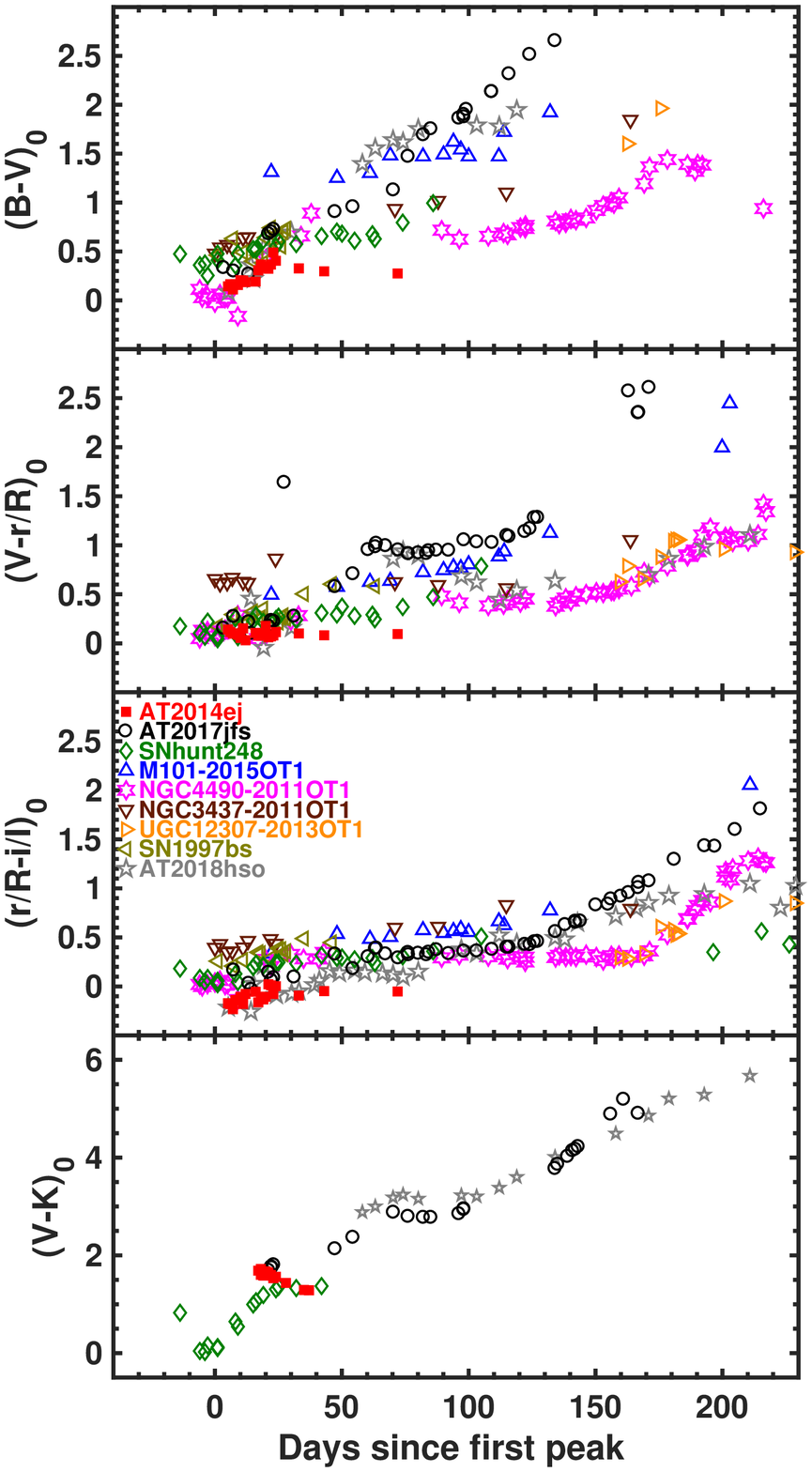}
\caption{Intrinsic broadband color evolution of AT~2014ej compared to other well-observed LRNe, including
SN~1997bs  \citep{2000PASP..112.1532V},
NGC~3437-2011-OT1 \citep{2019A&A...630A..75P}, 
NGC~4490-2011-OT1 \citep{2019A&A...630A..75P}, 
UGC~12307-2013-OT1 \citep{2019A&A...630A..75P},  
M101-2015-OT1 \citep{2017ApJ...834..107B}, 
SNhunt248 \citep{kankare_snhunt248}, 
AT~2017jfs \citep{2019A&A...625L...8P}, 
and AT~2018hso \citep{at2018hso}.
The colors plotted were corrected for both Milky Way and host-galaxy reddening using the values summarized in Table.~\ref{tab:comp_obj_at2014ej}.}  
\label{Fig:colorAT2014ej}
\end{figure}

In the three top panels of Fig.~\ref{Fig:colorAT2014ej} we plot the intrinsic optical colors of AT~2014ej compared to those of known LRNe candidates:
SN~1997bs  \citep{2000PASP..112.1532V,2019A&A...630A..75P},
NGC~3437-2011-OT1 \citep{2019A&A...630A..75P}, 
NGC~4490-2011-OT1 \citep{2019A&A...630A..75P}, 
UGC~12307-2013-OT1 \citep{2019A&A...630A..75P},  
M101-2015-OT1 \citep{2017ApJ...834..107B}, 
SNhunt248 \citep{kankare_snhunt248}, 
AT~2017jfs \citep{2019A&A...625L...8P}, and
AT~2018hso \citep{at2018hso}. The comparison objects have been corrected for reddening, adopting $E(B-V)_{\rm tot}$ color excess values listed in Table~\ref{tab:comp_obj_at2014ej}. 

Our  observations of AT~2014ej cover the first $+71$~d of color evolution corresponding to the object's first and second peak. At the earliest epochs, the intrinsic $(B-V)_{0}$ color is $\approx$0. This places it among the bluest objects in the comparison sample, similar to NGC~4490-2011-OT1. Over the period following the first light-curve peak, up to $+$24~d, the colors of AT~2014ej also evolve very similarly to those of the comparison sample. 
Close inspection reveals that  the  $(B-V)_{0}$ colors of AT~2014ej evolve rapidly to red, similar to the comparison sample. However, during the rise and after the second peak,
the $(B-V)_{0}$ and $(V-r)_{0}$ colors of AT~2014ej are clearly bluer than those of the other objects.

To conclude our investigation into the color curves of AT~2014ej, we plot  in the bottom panel of Fig.~\ref{Fig:colorAT2014ej} is its  intrinsic $(V-K)_0$ color curve along with those of  AT~2017jfs, SNhunt128, and AT~2018hso. Although the temporal coverage of AT~2014ej is relatively limited, its colors  match  those of the comparison objects well during the overlapping phases. This provides additional confidence that AT~2014ej is indeed an LRN.

\begin{figure}
\centering
\includegraphics[width=18cm]{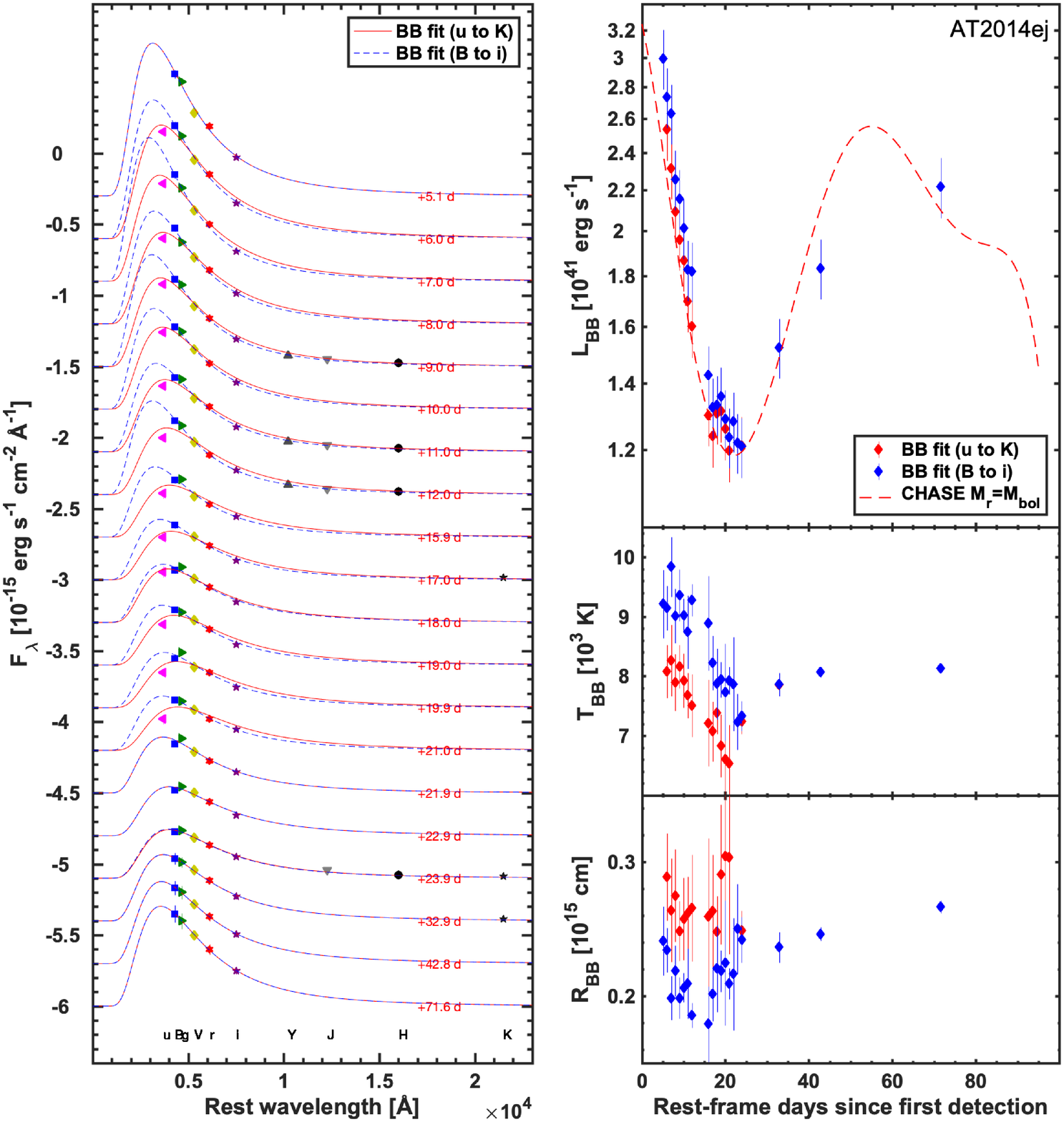}
\caption{\textit{Left:} SEDs of AT~2014ej constructed from photometry obtained between $+$5.1~d to $+$71.6~d post discovery. On each SED we overplot its best-fit BB function (solid red line) including all the available data from $u$ to $K_s$, and its best-fit BB function (dashed blue line) including only the filters in common for all the epochs, namely $B$ to $i$. \textit{Right:} Estimates of BB luminosities (from the BB fit integrations and from the CHASE $r$ band) (top),   BB temperatures (middle), and BB radii (bottom).} 
\label{Fig:BB2}
\end{figure}

We constructed the spectral energy distributions (SEDs) of AT~2014ej following the same procedure as for SNhunt120 (see Paper~1, Sect.~3.1). 
We plot in Fig.~\ref{Fig:BB2} the reddening-corrected SEDs of AT~2014ej.
These SEDs are relatively well reproduced by a single blackbody (BB) at all epochs (red solid line in the left panel), including the epochs when some NIR ($YJHK_s$) photometry is also available.
 Because of line blanketing, the $u$-band flux points lie below the fitted BB function.
 To facilitate a more consistent comparison of the bolometric properties at epochs with different wavelength coverage, we also present a single BB fit limited to the bands in common at all epochs, that is, from $B$ to $i$,  and this is plotted as dashed blue lines in the left panel. Because the $u$ band was excluded, we obtain BB curves that tend to peak at bluer wavelengths than when we include all the bands in our BB fitting.
 
\begin{figure}
\centering
\includegraphics[width=18cm]{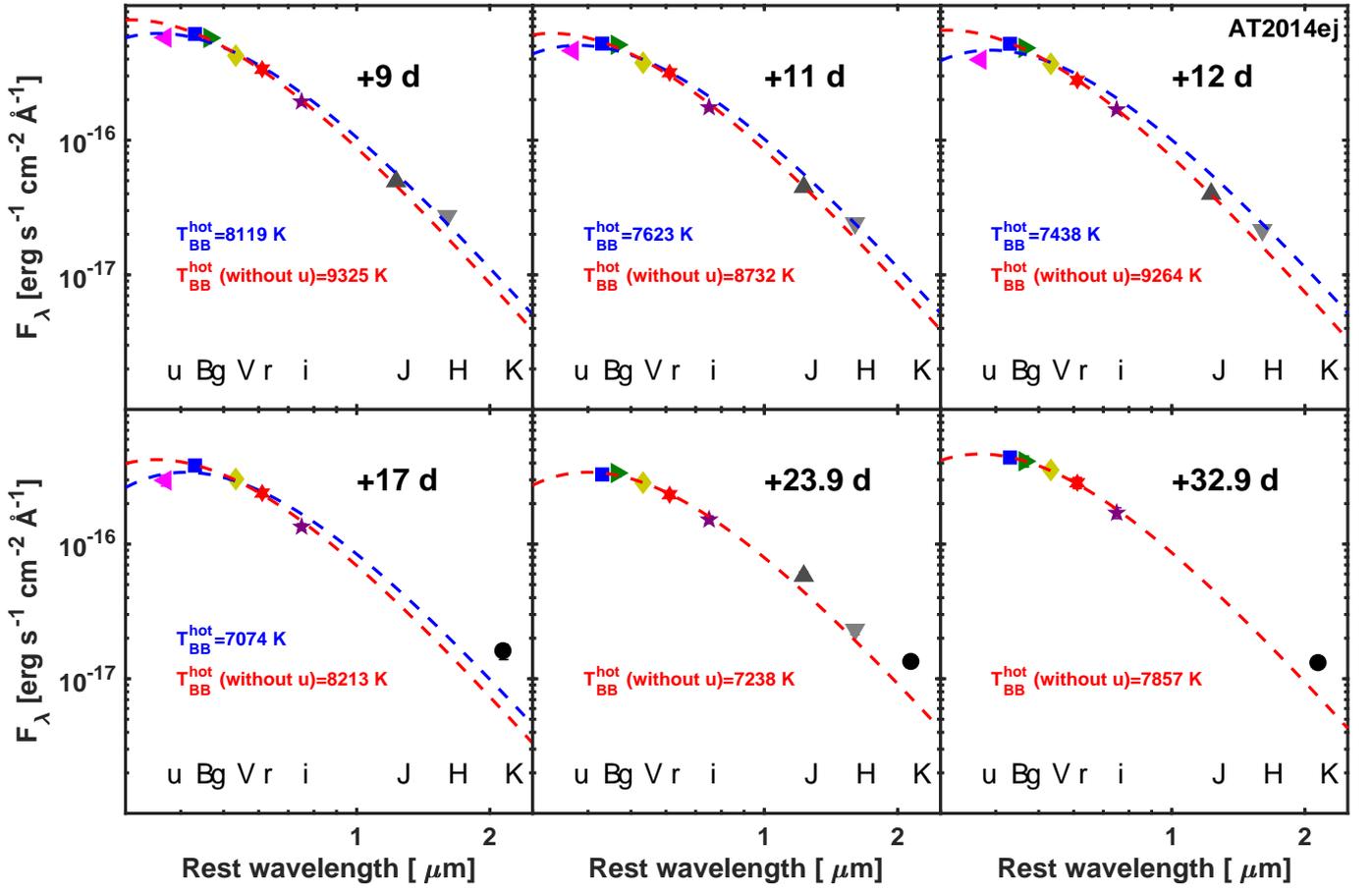}
\caption{SEDs of AT~2014ej, constructed with $uBgVri$- and $JHK_s$-band flux points for the six epochs where at least some NIR observations are available. The phase of each SED is reported in black in each panel. The flux points are fit with a single BB function including (dashed blue line) and excluding (dashed red line) the $u$-band flux measurement. The three epochs of $K_s$-band flux points  suggest an excess of flux relative to the best-fit single BB function, and this applies regardless of whether  the $u$-band flux point is included in the fit. This $K_s$-band excess could be indicative of  dust emission, but without MIR coverage, it is difficult to be conclusive.} 
\label{Fig:sed_kband}
\end{figure}

On $+$9~d, $+$11~d, $+$12~d, $+$17~d, $+$23.9~d, and $+$32.9~d, we have NIR broad-band measurements. When plotted along with the other flux points in log-log space, they reveal a non-negligible  excess of flux in the $K_s$ band, as compared to the single BB fit (including and excluding $u$ band), which might be an indication of dust  emission  at  redder wavelengths. 
These six SEDs are plotted in Fig.~\ref{Fig:sed_kband} along with single BB fits that include and exclude the $u$-band flux point. A conclusively determination whether the $K_s$ excess is due to dust emission would require mid-IR (MIR) observations, which unfortunately were not taken.

The BB-radius and BB-temperature profiles associated with the photosphere and the bolometric light curve of AT~2014ej are also plotted in Fig.~\ref{Fig:BB2}.
Profiles for both of these parameters are shown for fits including all the bands and fits only including  between the $B$ and $i$ bands.
Excluding the $u$ band produces a radius that is smaller by about 0.05$\times$10$^{15}$ cm than including it, and a higher temperature by about $\sim$1000~K. The BB luminosity does not change significantly between the two BB fit approaches. 
The bolometric properties obtained by fitting a BB over the wavelength range including the $B$ and $i$ bands show that
the BB radius initially declines from $\sim 0.25\times10^{15}$~cm to below $0.2\times10^{15}$~cm, and then, beginning  at about  $+$20~d, rises to $0.6\times10^{15}$~cm. The BB temperature declines from $\sim,10,000$~K to $\sim7000$~K, followed by an increase to 8000~K during the rise to and following the second peak.

After the initial (until +15~d) decline (from 3.0 to 1.3$\times10^{41}$ erg~s$^{-1}$), the luminosity remains almost constant, reaching $\sim$1.2$\times$10$^{41}$ erg~s$^{-1}$ at $+$24 days. 
The complete multiband CSP-II light-curve data of AT~2014ej only extend to +24~d. However, when we consider the CHASE $r$-band light curve, the PESSTO $V$- and $K_s$-band data, and  spectrophotometry, we can add a few points that show a rebrightening at +33~d, +43~d and +72~d, forming a second peak. 
When we convert the CHASE absolute $r$ band to a bolometric luminosity assuming $M_r \approx M_{Bol}$, the luminosity evolution of AT~2014ej is extended up to $+$95~d. This is overplotted in the top right panel of Fig.~\ref{Fig:BB2} as the dashed line. The $r$~band  reproduces the  bolometric light curve rather well after the initial epochs when the flux comes predominantly from  bluer wavelengths. If our assumption is valid, the luminosity has a second peak that reaches a luminosity of $2.6\times10^{41}$ erg~s$^{-1}$.

\subsection{Spectroscopy}
\label{AT2014ejspectroscopy}

\begin{figure}
\centering
\includegraphics[width=18cm]{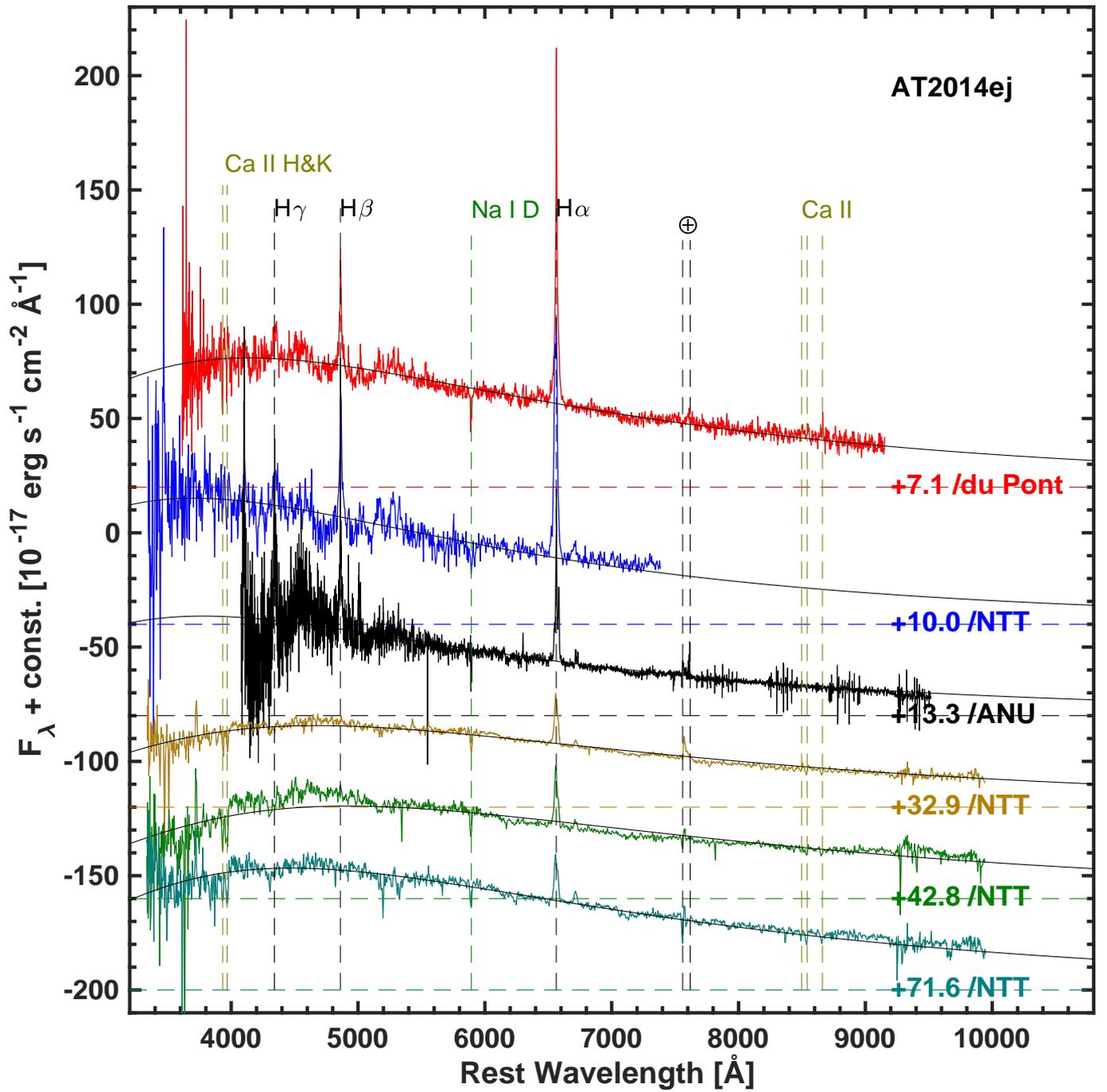}   
\caption{Low-resolution reddening-corrected visual wavelength spectra of AT~2014ej plotted in the rest-frame. The spectra are offset from one another in flux space by the addition of an arbitrary constant for presentation purposes. Epochs of observations  relative to the discovery date are provided, along with the labeling of key spectral features, and the positions of two prevalent telluric features are indicated with an Earth symbol. The solid black lines correspond to BB fits. 
The BB fits exclude the wavelength ranges that contain  strong emission lines. }
   \label{Fig:AT2014ejspectroscopy}%
    \end{figure}

\begin{figure}
\centering
\includegraphics[width=18cm]{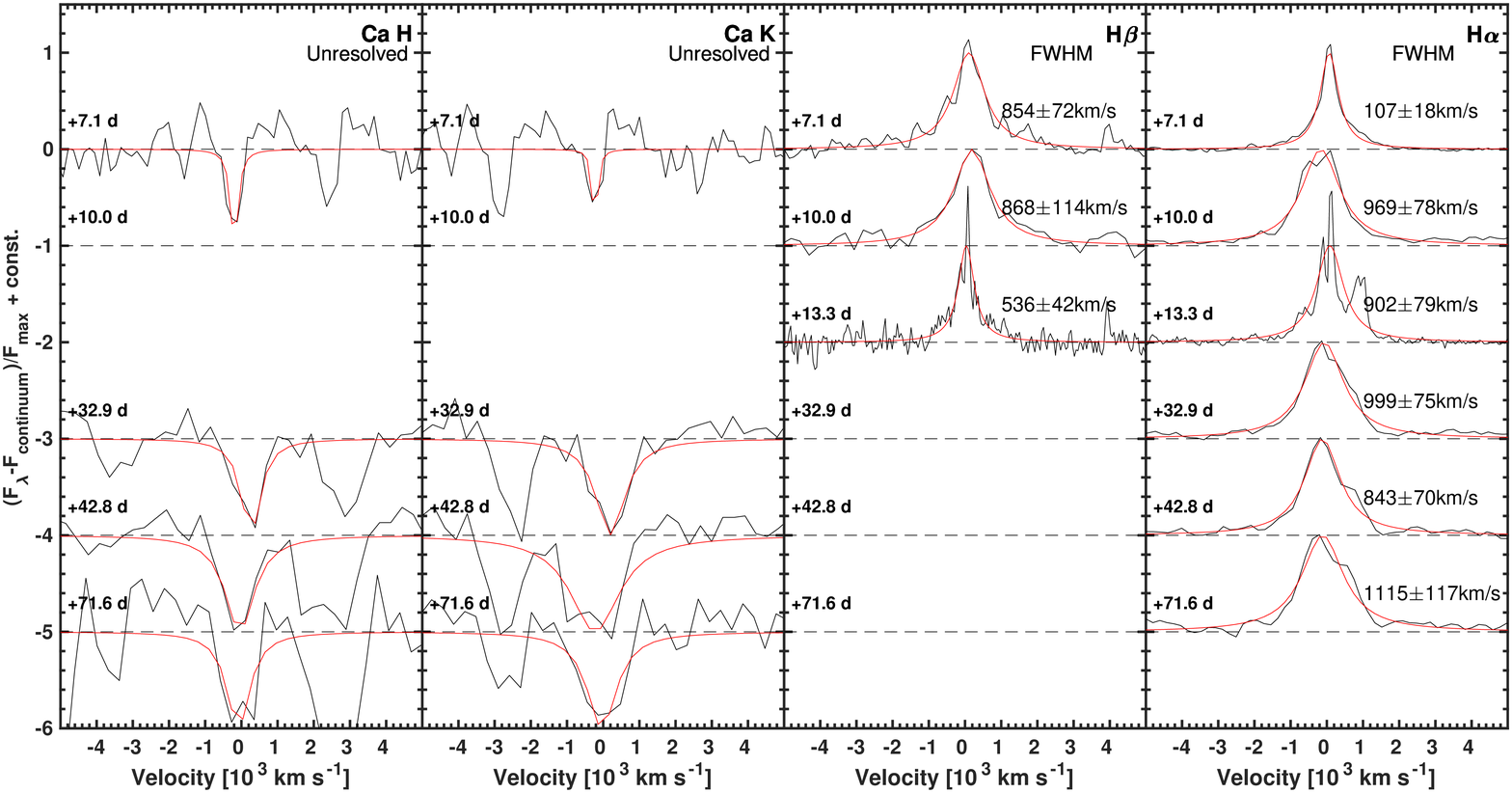}  

\includegraphics[width=9cm]{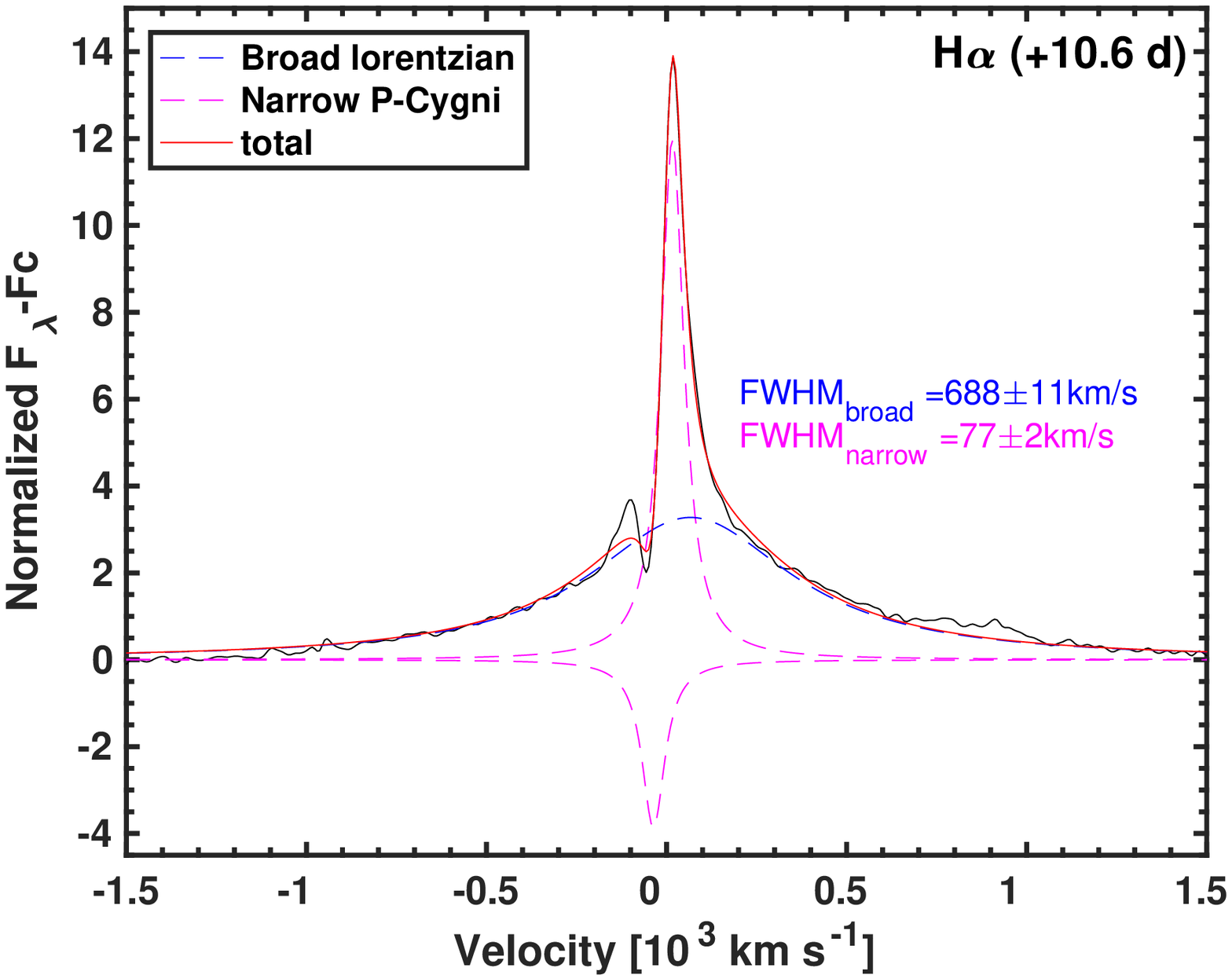}   
  \caption{\textit{Top:} spectral features of AT~2014ej including the \ion{Ca}{ii} H\&K, H$\beta$, and H$\alpha$ lines.  The  continuum of each spectrum was fit  using a low-order polynomial and was subtracted from the  data.
 The Lorentzian line profile fits are overplotted with a red line, and the corresponding line velocities are listed and also provided in Table~\ref{tab:FWHM_AT2014ej}.
 \textit{Bottom:} MIKE spectrum around H$\alpha$ in velocity space. The profile is well fit by two components consisting of  a broad Lorentzian and a narrow P~Cygni profile (sum of two Lorentzians with the same width) whose FWHMs are reported in the figure.}
\label{fig:lineprofilesAT2014ej}
    \end{figure}

Six epochs of low-resolution visual wavelength spectra  of AT~2014ej are plotted in Fig.~\ref{Fig:AT2014ejspectroscopy}; they extend between +7~d and +72~d relative to first detection.
The main features characterizing the spectra are the narrow Balmer lines in emission and \ion{Ca}{ii}~H\&K in absorption.
The spectral continuum becomes redder with time.
 The velocities of the lines, corrected for the spectral resolution, are shown in the top panel of Fig.~\ref{fig:lineprofilesAT2014ej}, where the Balmer lines are shown to have a full width at half-maximum (FWHM) of about 900 km~s$^{-1}$, while the \ion{Ca}{ii} absorption exhibits lower velocities and is unresolved.
 In the ANU (+ WiFes) spectrum at +13.3 d, which has a higher resolution than the NTT (+ EFOSC2) spectra, it is possible to see that the
 presence of [\ion{N}{ii}] from the host galaxy modifies the H$\alpha$ profile. The spectrum also indicates a narrow P~Cygni profile for H$\alpha$ and H$\beta$.
 
\begin{figure}
   \centering
   \includegraphics[width=18cm]{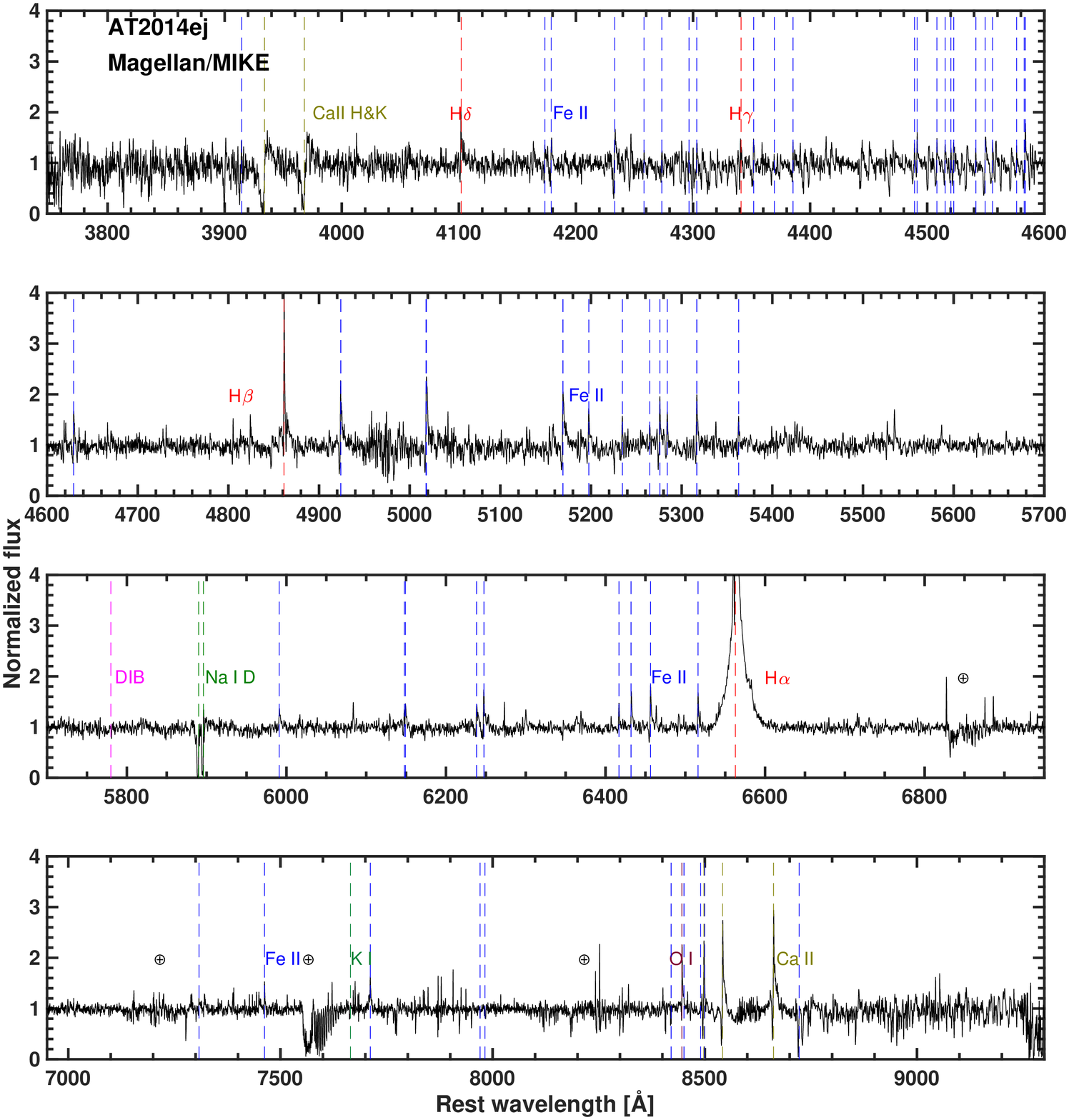}   
   \caption{Medium-resolution visual-wavelength spectrum of AT~2014ej obtained with the Magellan Clay (+ MIKE) telescope on $+$11~d and plotted in the rest-frame. Key spectral features are labeled, and  prominent telluric features are marked with an Earth symbol. }
   \label{Fig:AT2014ejhighres}
    \end{figure}

\begin{figure}
\centering
\includegraphics[width=18cm]{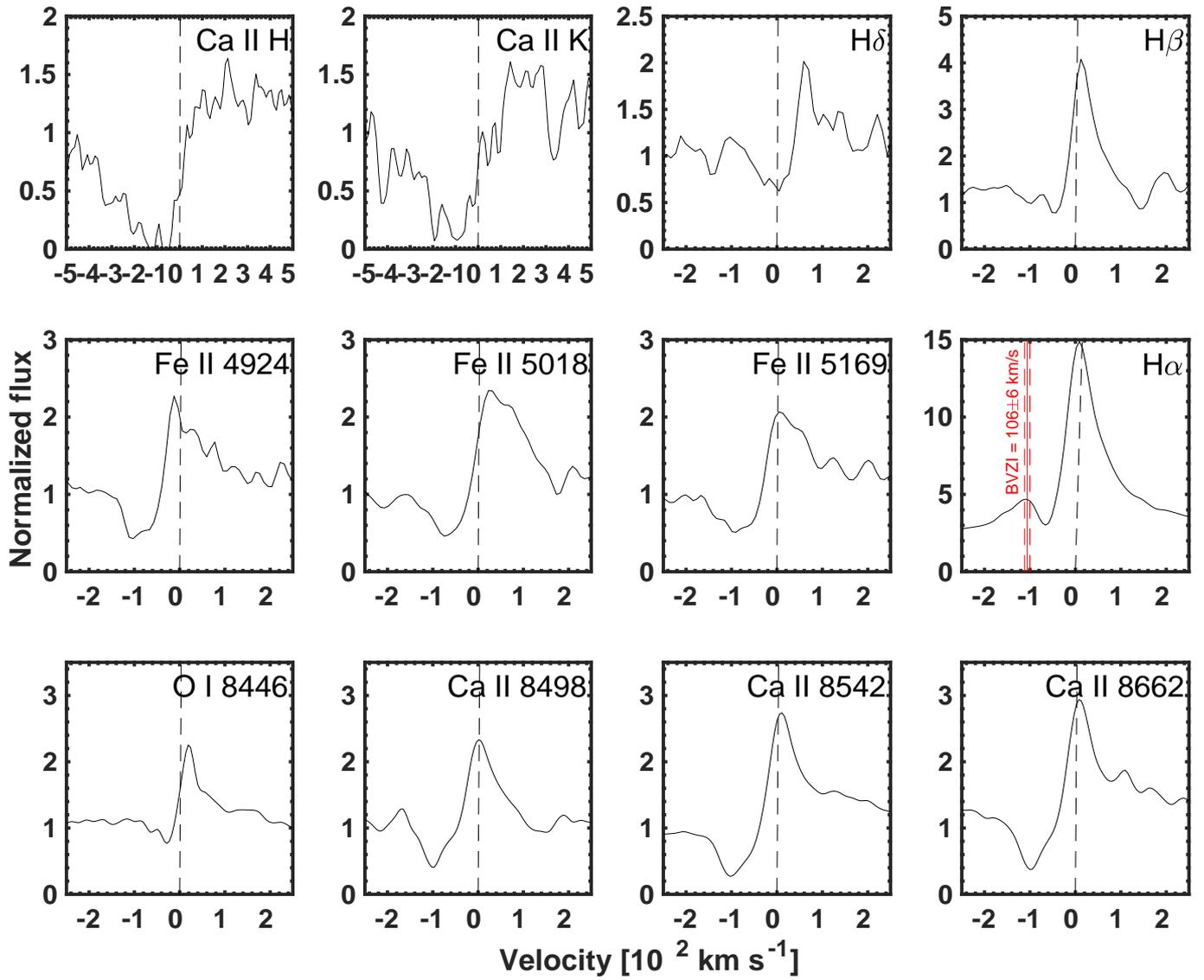} 
\caption{AT~2014ej line profiles of prominent emission lines in the medium-resolution MIKE spectrum  (see Fig.~\ref{Fig:AT2014ejhighres}), plotted in velocity space. This includes \ion{Ca}{ii} H\&K, Balmer lines of H$\delta$, H$\beta,$ and H$\alpha$, the \ion{Fe}{ii} multiplet 42, \ion{O}{i}~$\lambda$8446, and the \ion{Ca}{ii} NIR triplet. The vertical dotted lines mark  the rest-velocity positions of the ions. The red line in the $H\alpha$ panel marks the BVZI of $106\pm6$~km~s$^{-1}$.
}
\label{fig:lineprofilesNGC7552}%
 \end{figure}

The higher dispersion MIKE spectrum obtained on $+11$~d is shown in Fig.~\ref{Fig:AT2014ejhighres} and reveals a clear narrow P~Cygni component associated with each of the spectral features that are plotted (in velocity space) in  Fig.~\ref{fig:lineprofilesNGC7552}.
The narrow H$\alpha$ P~Cygni feature is measured to have a blue velocity at zero intensity (BVZI) of $106\pm6$~km~s$^{-1}$ and a P~Cygni minimum velocity of $66\pm6$~km~s$^{-1}$. 
This line component is observed in many SNe~IIn and is thought to be formed from outer unshocked circumstellar material (CSM) that is excited by the emission originating from the shock region \citep[see, e.g.,][]{kiewe12IIn,taddia13IIn,nyholm19}. As a result, the BVZI value of this component provides a measure of the CSM velocity. 

In the bottom panel of Fig.~\ref{fig:lineprofilesAT2014ej} we show the entire profile of the H$\alpha$ line in the MIKE spectrum. This is well reproduced by two components, a narrow P~Cygni (given by two Lorentzians, one in emission and one in absorption) and a broader Lorenzian emission. The broad component has a velocity of about 690~km~s$^{-1}$ , whereas the narrow one has an FWHM of about 80~km~s$^{-1}$. These two components are not visible in the spectra with lower resolution (with the possible exception of the ANU spectrum) that are shown in the top panel of Fig.~\ref{fig:lineprofilesAT2014ej}.
The broader Lorentzian component in the H lines shown here could form from electron scattering, which would alter the narrow emission profile to a broader profile in a region in close proximity to the location of the CSM interaction. 

The BVZI and absorption minimum velocities are approximately 100~km~s$^{-1}$ for all the lines except for \ion{Ca}{ii} H\&K, where the BVZI is about 500~km~s$^{-1}$. The higher velocity  inferred from the  \ion{Ca}{ii} H\&K feature  might arise because it originates from  material that is ejected during a massive binary merger.

\subsection{Progenitor constraints}
\label{AT2014ejprogenitorconstraints}

NGC~7552 was observed on a number of occasions with the HST (+ WFPC2) prior to
the discovery of AT~2014ej. Unfortunately, the location of AT~2014ej lies a
few pixels outside the field of view of WFPC2 in the earliest images taken in 2015 April. 
However, 2$\times$600s F218W images were obtained with HST (+ WFC3)  on 1999 March 31 that do cover 
the position of AT~2014ej.  
Here only the center of the galaxy was clearly visible in the F218W data, and the lack of reference sources in the field prevents us from obtaining  any precise astrometry. We
therefore performed photometry with {\sc hstphot} on a region
contained within a 40$\times$40 pixel box centered on the coordinates of the transient, and extracted all sources detected with a $>5$-sigma significance. Within this region lies no source.
With no candidate source in the region of the transient, we set an upper $F218W>20.3$ mag limit determined from the average magnitude of all point sources in the field with a $5$-sigma detection.


A further set of pre-outburst images covering the site of AT~2014ej was taken between 2009 May 5-6 with HST (+ WFPC2). This includes deep images with the broadband filters
$F336W$ (4400s), $F439W$ (1800s), $F555W$ (520s), and $F814W$ (520s), and as well with a narrow-band
H$\alpha$ (1800s) filter. The field of view (FOV) afforded by HST (+ WFPC2) is smaller than that of our post-outburst ground-based CSP-II data, and it is challenging to align the two images  to precisely localize  the position of AT~2014ej in the HST
images. Using five sources common to both images, we find that the position of AT~2014ej lies at 
 x,y pixel coordinate (75, 172) on the $F555W$ image.

\begin{figure}
   \centering
      \includegraphics[width=15cm]{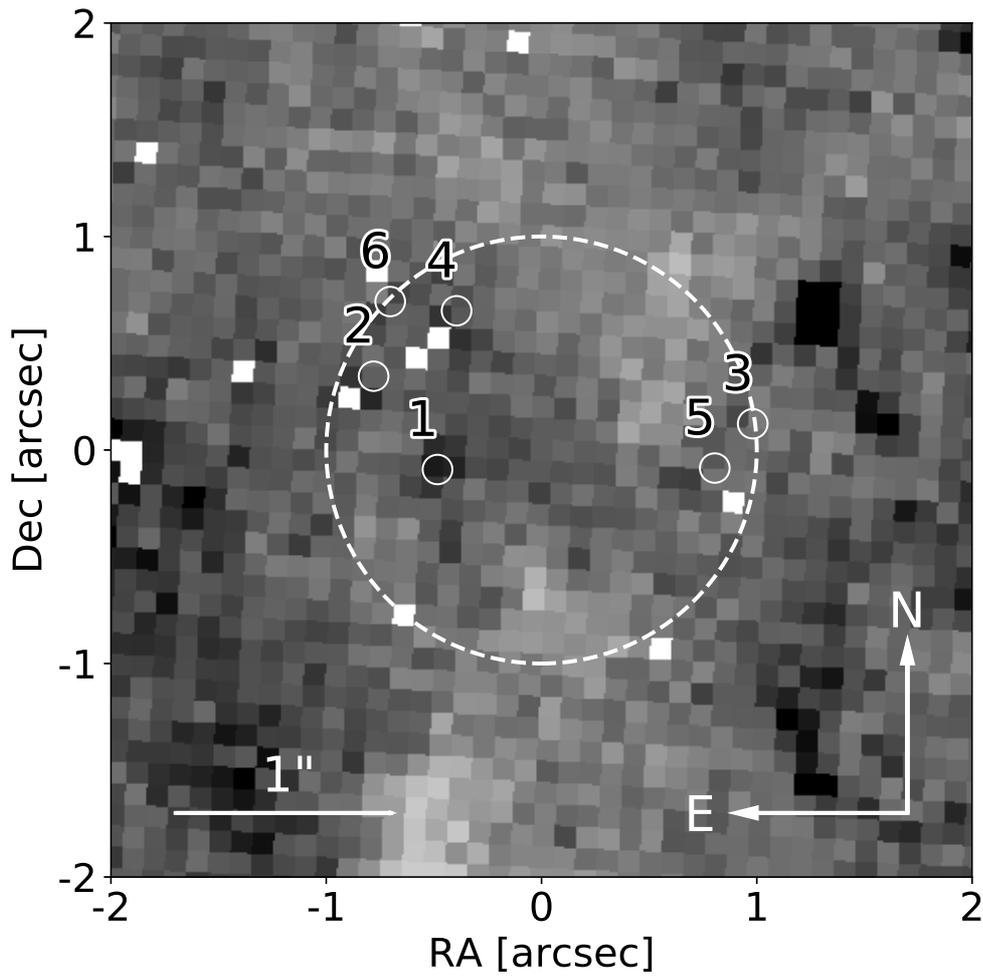}
  \caption{Archival HST (+ WFPC2) F555W-band image of NGC~7552. The position of AT~2014ej is indicated with a circle, and the positions of sources detected within 1\arcsec\ are numbered. Because most sources detected in this region have a low S/N, they are not particularly obvious on visual inspection. White squares are aberrant pixels that have been masked.}
    \label{Fig:progenitor2}%
   \end{figure}

While the formal uncertainty in the geometric transformation between pre- and post-outburst images is $<1$ WFPC2 pixel ($=0\farcs1$), this probably is an underestimate due to the small number of sources used for the alignment.  Adopting a conservative
approach,  we computed photometry for all sources detected at 
a  $>5$-sigma significance within 1\arcsec of the position of AT~2014ej, which might possibly be   progenitor candidates. None of the sources are measured to be brighter than the apparent magnitude $m_{F814W} \sim 23$. These sources are indicated in Fig.~\ref{Fig:progenitor2}, along with the location of AT~2014ej.  Here the area enclosed within the circle  accounts for the uncertainty in the estimated position of AT~2014ej.

\section{Discussion}
\subsection{AT~2014ej compared with luminous red nova candidates}

Here the similarities between the observational properties of AT~2014ej to those of other LRNe are assessed. 
First we present a spectral comparison between AT~2014ej and the well-observed LRN AT~2017jfs, and then we compare the light curves of AT~2014ej to AT~2017jfs and an expanded sample of LRNe from the literature. 

Visual wavelength spectra of AT~2014ej and AT~2017jfs are plotted in Fig.~\ref{Fig:AT2014ej_spectracomparison}. At early times, which coincide with their first peak (see below), the two objects are strikingly similar, exhibiting nearly identical narrow spectral features, most notably, narrow Balmer emission lines and some  metal lines   blueward of H$\beta$.  As the two objects evolve toward their second peak (see below), the strength of  H$\alpha$  decreases while the H$\beta$ feature  disappears.
Furthermore, beyond $+$60~d, the H$\alpha$ continues to decrease in AT~2014ej and becomes absent in AT~2017jfs. During the same period, the spectra of AT~2017jfs evolve further to the red, while those of AT~2014ej on $+$71.6~d show little evolution. 
This behavior may be linked to AT~2014ej being  brighter than AT~2017jfs at these phases (see below). Unfortunately, we lack spectral observations of AT~2014ej at even later phases, and are therefore unable to document the expected emergence of molecular overtones consistent with the spectrum of a late-type star and the restrengthening of H$\alpha$, as seen in LRNe such as AT~2017jfs  \citep{2019A&A...625L...8P}. 

In Fig.~\ref{Fig:abmagat2014ej} the absolute magnitude light curves of AT~2014ej are compared to those of eight other LRN candidates, including: SN~1997bs, NGC~3437-2011-OT1, NGC~4490-2011-OT1, UGC~12307-2013-OT1, M101-2015-OT1, SNhunt248, AT~2017jfs, and AT~2018hso. The comparison light curves have been adjusted for reddening  
and placed on the absolute magnitude scale, for which we adopted the color excess values and the distance moduli listed in Table~\ref{tab:comp_obj_at2014ej}.  
Although the light curve of AT~2014ej is limited in phase coverage, its shape and absolute magnitudes resemble the double peak of the other well-observed LRNe.
In particular, the $V$ and $r$ bands evolve very much like that of AT~2017jfs.  
The CHASE unfiltered light curve of AT~2014ej reveals a second peak that is $\lesssim$0.23 mag fainter than at discovery. The strength and phase of the secondary peak of AT~2014ej is similar to that of SNhunt248; it  occurs about 100~d earlier than the peaks of M101-2015-OT1 and NGC~4490-2011-OT1, both of which exhibit a second peak that occurs well after $+$100~d. 

We conclude this part of the analysis with a return to Fig.~1 of Paper~1, which compares the peak absolute $r$-band magnitude with the $r$-band light curve decay (defined as the time in days  for the light curve to drop one magnitude). Our lower limit value of AT~2014ej places it well within the  parameter space that is populated by other well-observed LRNe. When we consider that AT~2014ej was discovered somewhat past its first peak,   it is clearly one of the brightest LRN studied so far (see below).

\subsection{Direct observational evidence for linking (luminous) red novae to common-envelope events}

Because a common-envelope (CE) phase is short \citep[about $10^3$~yr,][]{mmh79,pod01}, it is extremely rare to detect this ephemeral phase of stellar evolution. Nevertheless, a number of candidates for direct observations of CE events have been proposed, with  a prime candidate being red novae. In particular, V1309~Sco (a V838~Mon-class event) serves as one of the most promising examples of an active CE event (or merger) that is caught in action \citep []{tylenda11,ijc+13}.
V1309~Sco was the first stellar event to provide conclusive evidence that contact binary systems end their evolution in a stellar merger.
The rate of red novae events has been estimated to be as high as 20\% of the core collapse SN rate \citep{thompson09}. However, this 20\% was computed without considering the difference between LRNe and ILRTs, which means that they were all considered to be the same type of objects. This value was therefore overestimated. A value of $>$5\% was discussed in \citet{2009MNRAS.398.1041B} when they presented SN 2008S.

Another promising example with a convincing link to a CE event is that of the LRN M31-2015-OT1 , whose final progenitor binary system has been investigated \citep{wdbs15,2017MNRAS.470.2339L,Blagorodnova2020} and simulated in detail \citep{mmr+17}. It has been suggested that this event, along with a number of other LRNe,
might simply be more scaled-up versions of normal red novae (typically fainter than $M_V \sim -10$), thus resulting from merger events involving more massive stars than the less luminous red novae \citep{2017ApJ...834..107B,2016MNRAS.458..950S,2017MNRAS.470.2339L,2018MNRAS.473.3765M,mos18,2019A&A...630A..75P}. 
There seems to be consensus in the literature for a direct link between the luminosities of various red novae phenomena and the masses of the stellar components undergoing coalescence (CE evolution).
More specifically, the fainter red novae involve stellar masses of $\sim 1-5\;M_\odot$, such as V838~Mon and most likely have $5-10\;M_\odot$ components; whereas the LRNe stellar masses are thought to be a few $10\;M_\odot$ \citep{2017MNRAS.471.3200M}. 
In a recent population synthesis study, \citet{hsv+19} reported a Galactic LRNe rate of $\sim 0.2\;{\rm yr}^{-1}$, in agreement with the observed rate. They also argued that the Large Synoptic Survey Telescope will observe 20--750~LRNe per year. Such a rate  will enable us to estimate the luminosity function of LRNe and reveal the diversity  among CE events.

\citet{2019A&A...630A..75P} provided a comprehensive review and discussion of the properties and progenitors of red novae and LRNe.
The double-peaked light curves are still debated, but could be explained with a CE event by an initial ejected envelope, producing the low-luminosity light curve peak, followed by the merger of the secondary star onto the core component of the primary star. 
\citet{2017MNRAS.471.3200M} proposed that these double-peaked light curves arise from a collision between a dynamically ejected fast shell of material associated with the terminal stage of the CE merger with preexisting slow equatorial circumbinary material shed from the plunge-in of the secondary star during the early phase of in-spiral. In this scenario, the first optical peak arises through cooling envelope emission, and the second light-curve peak 
is powered by subsequent radiative shocks in the equatorial plane.  Moreover, differences between various red gap transients might be explained by different viewing angles of observers with respect to the orbital plane of the progenitor binary \citep[see Fig.~2 of][]{2017MNRAS.471.3200M}, along with several other key parameters characterizing the progenitor system. 

\citet{ggpp13} have studied the evolution of stellar mergers formed by a collision involving massive stars. They concluded that mass loss from the merger event is generally small ($<10$\% of the total mass for equal-mass star mergers at the end of the main sequence) and that little
hydrogen is mixed into the core of the merger product.  
This amount of mass loss may be sufficient to explain the double-peaked light curves in LRNe events, making this a viable model.  

\subsection{UVOIR light curves of luminous red novae}
 
 We plot in Fig.~\ref{Fig:bolocomp14ej} the UVOIR light curve of AT~2014ej, compared to those of our  LRNe comparison sample.    The overall morphology of the entire sample is similar,  
 with all of the objects displaying a double-hump light curve typically of most  LRNe. 
  However, as previously noted by \citet{2019A&A...630A..75P} and as shown in Fig.~\ref{Fig:bolocomp14ej}, there is diversity. This includes  the  luminosity  of  both peaks and the phase of the second peak relative to the first peak.  Nevertheless, in general, there appears to be a correlation between the luminosity of the two peaks. In other words, objects whose  first peak is located at the  bright end of the luminosity distribution (i.e., AT~2017jfs, SNhunt248, NGC~4490-2011-OT, and AT~2014ej) also have secondary maxima at the bright end of the luminosity distribution. If this trend holds for a more populated sample, we might speculate that the luminosity of the second peak may scale with the masses of the involved ejecta and CSM  undergoing circumstellar interaction.

  We also plot in  Fig.~\ref{Fig:bolocomp14ej} the two closest matching   light-curve  models published by \citet{2017MNRAS.471.3200M} and  discussed above.
  These include their fiducial  model computed for a 10  $M_{\odot}$ system characterized with the semimajor axis $a=30 R_{\odot}$ and with a predynamical mass runaway time of $t_{run} = 10\cdot t_{dyn}$, where $t_{dyn}$ corresponds to the in-spiral dynamical timescale. 
  The second synthetic light-curve model corresponds to   the fiducial model  with the mass increased to 30~$M_{\odot}$. This results in a brighter light curve relative to the fiducial model, but in general, the model fails to reach the luminosities exhibited at both peaks (by a factor of $\sim$10-15) for all but the faintest LRN M101-2015-OT1.  In light of the differences in luminosity and the timescales describing the rise and decline of the second peak, it would be interesting to see  the model calculations extended to even more massive systems, higher values of $t_{run}$, and even modifications to the  assumed  growth rate of predynamical mass loss.

\section{Conclusion}
\label{sec:conclusions}

In this study we have added observations to the growing populations of luminous red novae (LRNe). 
Our observational data set of AT~2014ej   demonstrates  characteristics 
of other LRNe, including the  double-humped
 light-curve evolution that is ubiquitous to this class of red transients.  
These double-humped light curves are reminiscent of the long-lasting type~IIn SN~2006jd, which \citet{2012ApJ...756..173S} argued to be produced by SN ejecta interacting with CSM located in a  torus around the progenitor, and ejected during its pre-SN phase \citep{1994MNRAS.268..173C}.   
Similarly, \citet{2017MNRAS.471.3200M}  attributed the  second peaks  observed in  most LRN to material ejected following  the CE merger of two massive stars that interacts with slow equatorial circumbinary material, previously ejected during the early phase of in-spiral. 
The \citeauthor{2017MNRAS.471.3200M} models plotted in Fig.~\ref{Fig:bolocomp14ej} exhibit a  similar trend in their overall evolution to the trend in the LRN comparison sample.
However, the models  underpredict the  luminosities
 of both peaks, and the timescales describing the evolution of the second peak differ. 
 Perhaps future efforts of modeling such systems should extend beyond the parameter space thus far considered. 

To conclude, we wish to consider that less than a decade ago, our understanding of gap transients was significantly limited.  In just a short time period, however, the  landscape of gap transients has grown to be populated by various subtypes ranging from  LBV eruptions to ILRTs possibly associated with electron capture supernovae,   LRNe produced during the CE phase of massive binaries, and  even to the faintest 2008ha-like SNe  whose origins remain unknown.  
 Efforts to further unravel the intricacies of gap transients will require detailed followup over long periods, while significant advancement in identifying the underlying progenitor populations of these objects can be expected with IR observations made  with the James Webb Space Telescope. 

\begin{acknowledgements}
We thank the referee for a constructive report that improved the presentation of our manuscript.
A special thanks to  Ian Thompson for obtaining the medium-dispersion spectrum of AT~2014ej and  to Massimo Della Valle for endearing comments.
The CSP-II has been funded by the USA's NSF under grants AST-0306969, AST-0607438, AST-1008343, AST-1613426, AST-1613455, AST-1613472 and also in part by  a Sapere Aude Level 2 grant funded by the Danish Agency for Science and Technology and Innovation  (PI M.S.).
M.S., F.T. and E.K. are supported by a project grant (8021-00170B) from the Independent Research Fund Denmark (IRFD). Furthermore, M.S. and  S.H. are supported  in part by a generous grant (13261) from VILLUM FONDEN. 
M.F. is supported by a Royal Society - Science Foundation Ireland University Research Fellowship.
T.M.T.\ acknowledges an AIAS--COFUND Senior Fellowship funded by the European Union’s Horizon~2020 Research and Innovation Programme (grant agreement no~754513) and Aarhus University Research Foundation. 
L.G. is funded by the European Union's Horizon 2020 research and innovation programme under the Marie Sk\l{}odowska-Curie grant agreement No. 839090. This work has been partially supported by the Spanish grant PGC2018-095317-B-C21 within the European Funds for Regional Development (FEDER).
Support for J.L.P and G.P.  is provided by the Ministry of Economy, Development, and Tourism’s Millennium Science Initiative through grant IC120009, awarded to The Millennium Institute of Astrophysics, MAS.
N.B.S. acknowledges support from the NSF through grant AST-1613455, and through the Texas A\&M University Mitchell/Heep/Munnerlyn Chair in Observational Astronomy.
Support for J.L.P. is provided in part by FONDECYT through the grant 1191038 and by the Ministry of Economy, Development, and Tourism’s Millennium Science Initiative through grant IC120009, awarded to The Millennium Institute of Astrophysics, MAS.
This research has made use of the NASA/IPAC Extragalactic Database (NED), which is operated by the Jet Propulsion Laboratory, California Institute of Technology, under contract with the National Aeronautics and Space Administration. We acknowledge the usage of the HyperLeda database.
Based on observations made with ESO Telescopes at the La Silla Paranal Observatory
under programmes 191.D-0935 and 096.B-0230.
\end{acknowledgements}

  \bibliographystyle{aa} 

%
%
%
%
%
%
%

\clearpage






 \clearpage

\clearpage

\clearpage

\clearpage
    
\clearpage
    
\clearpage

    \clearpage

\clearpage

 \clearpage
 
 \clearpage
   \begin{figure}
   \centering
      \includegraphics[width=18cm]{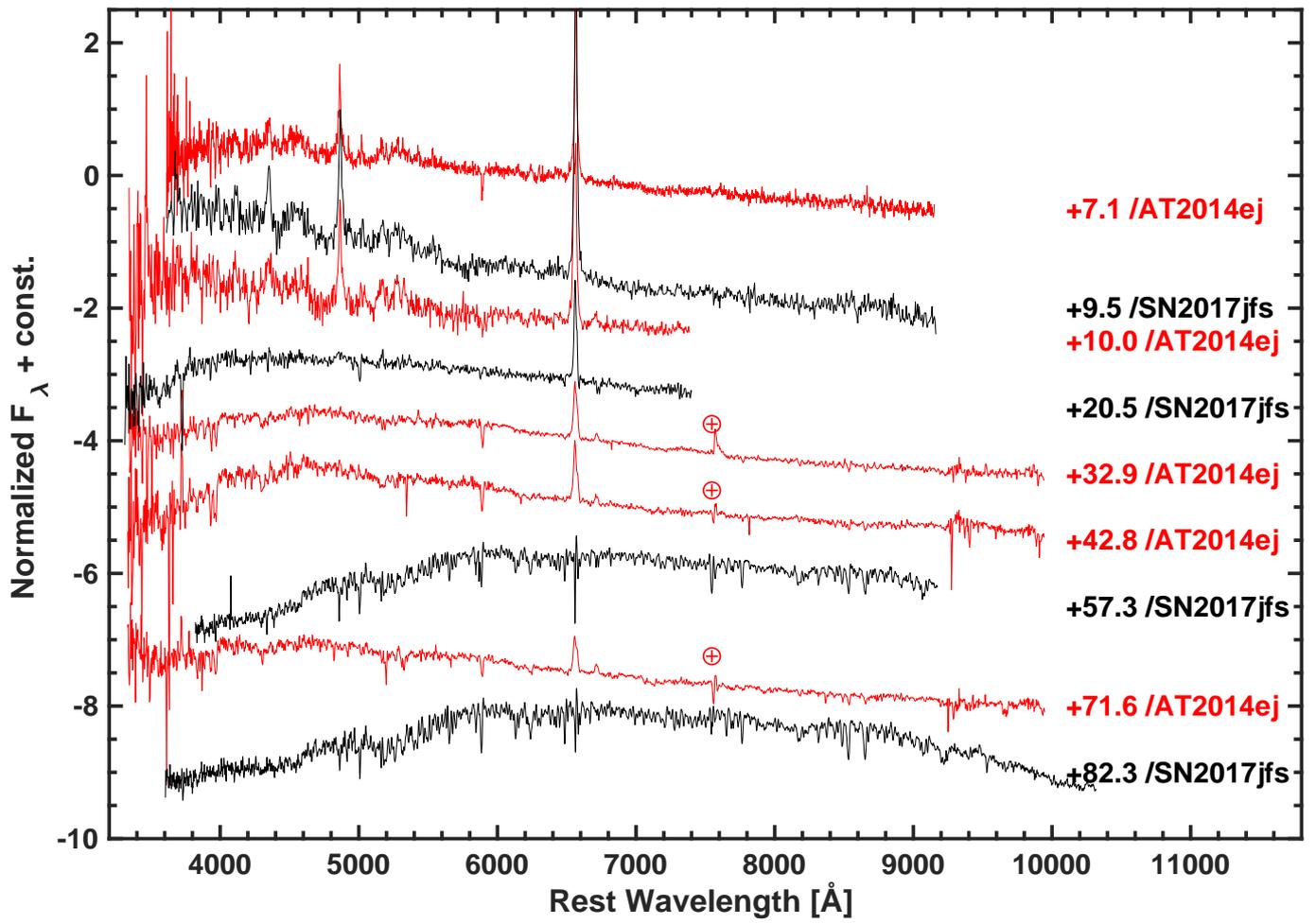}
   \caption{Comparison of  visual wavelength spectra of AT~2014ej with similar-epoch spectra of the LRN AT~2017jfs. Each of the spectra was corrected for reddening and is plotted in the rest-frame. Reported phases are in days relative to the first peak.}
   \label{Fig:AT2014ej_spectracomparison}
   \end{figure}

\clearpage
\begin{figure}
\centering
\includegraphics[width=18cm]{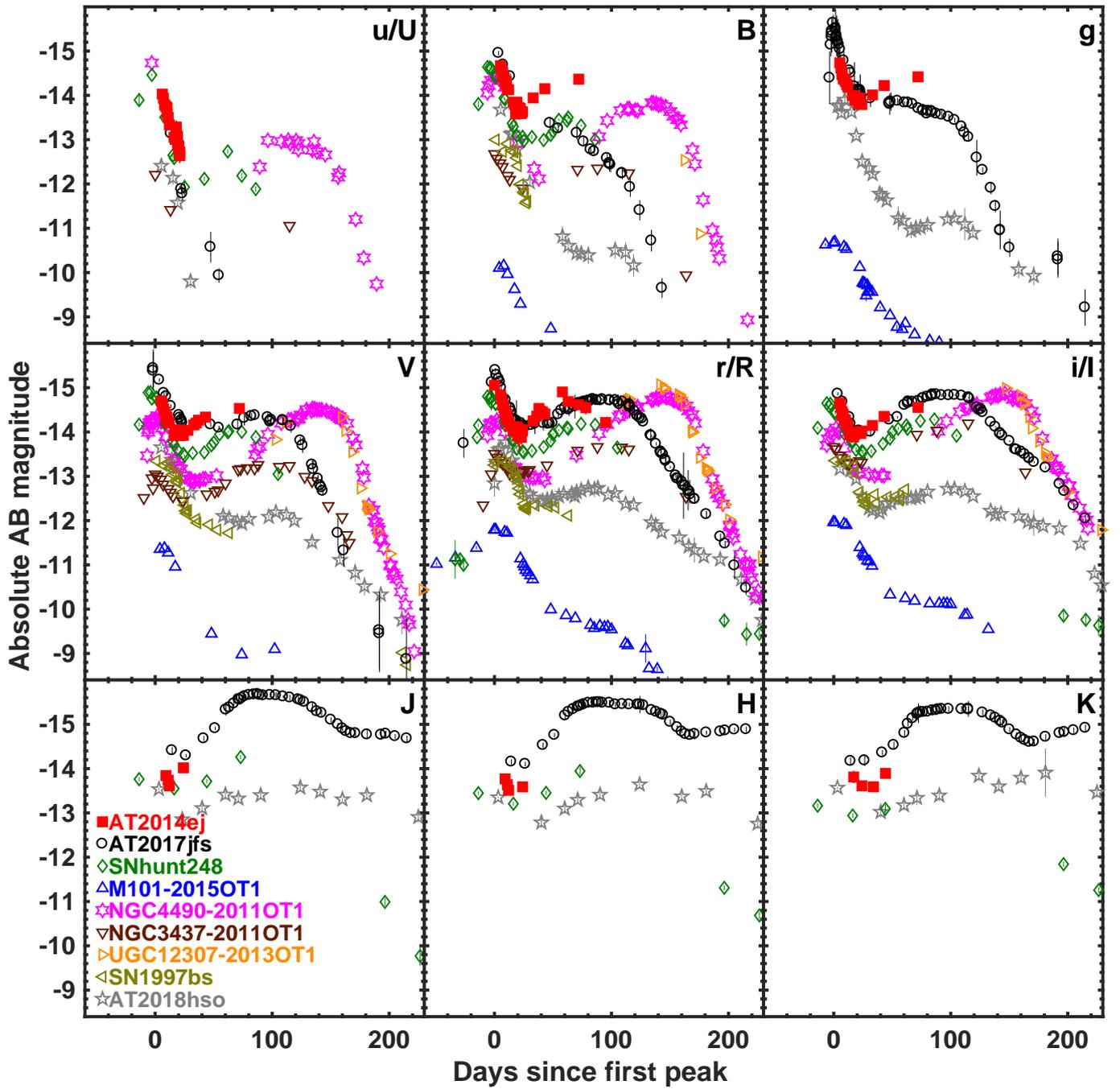}
\caption{Absolute magnitude light curves of AT~2014ej compared to other well-observed LRNe: SN~1997bs, NGC~3437-2011-OT1, NGC~4490-2011-OT1, UGC~12307-2013-OT1,  M101-2015-OT1, SNhunt248, AT~2017jfs, 
and AT~2018hso.
Each  light curve is on the AB system.}
\label{Fig:abmagat2014ej}
\end{figure}

\clearpage
\begin{figure}
\centering
\includegraphics[width=18cm]{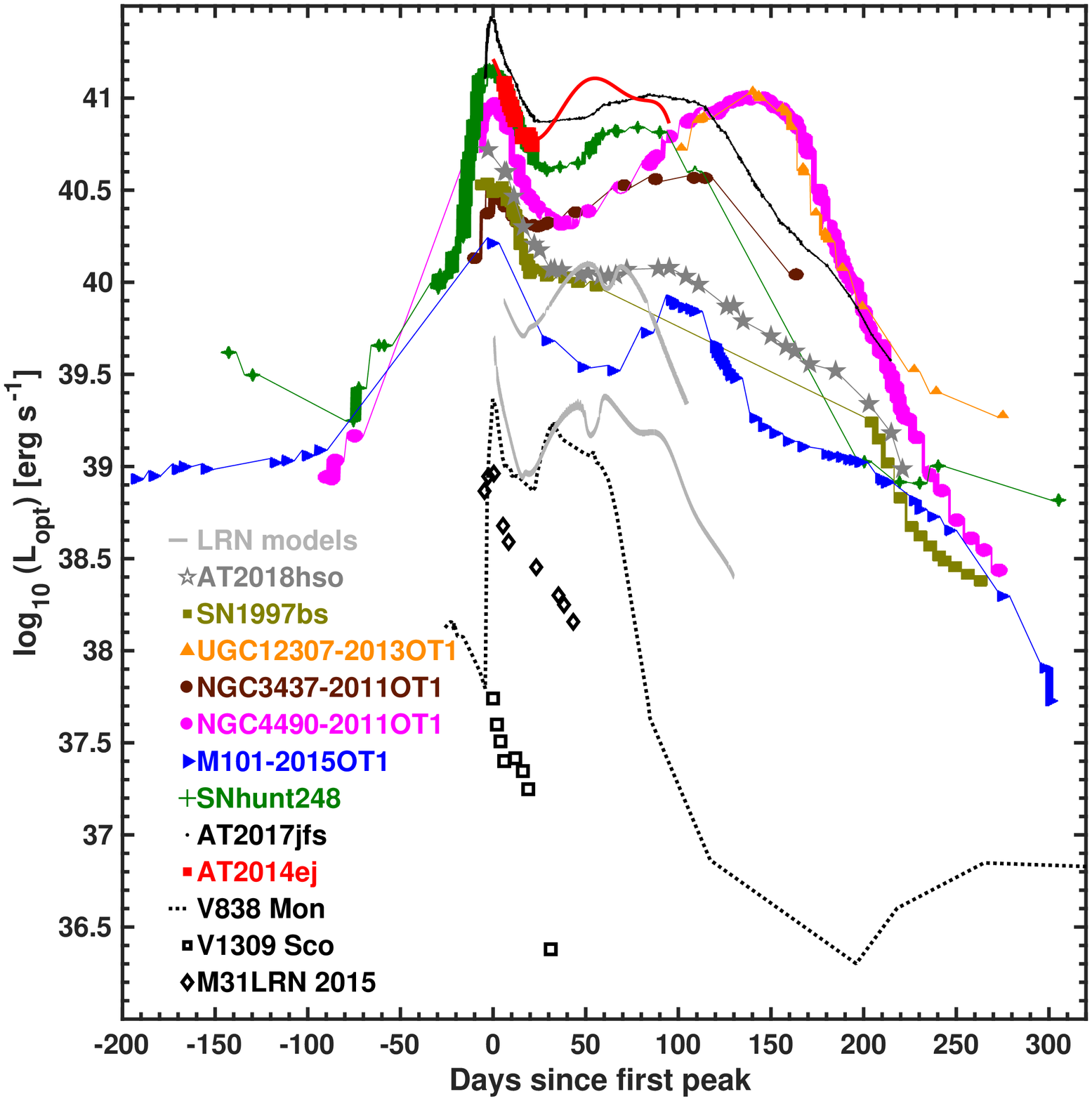}
\caption{Pseudo-bolometric light curves of a handful of well-observed LRNe from  \citet{tylenda05}, \citet{tylenda11}, \citet{wdbs15}, \citet{at2018hso}, and \citet{2019A&A...630A..75P}, plotted along with that of AT~2014ej. The solid red line extended beyond the UVOIR light curve of AT~2014ej corresponds to CHASE unfiltered photometry.  Two LRN models published by \citet{2017MNRAS.471.3200M} are also overplotted as gray lines. This includes the light curve computed for  their fiducial model assuming  a binary mass of  10 $M_{\odot}$, a semimajor axis of $a=30 R_{\odot}$, and a predynamical mass runaway time of $t_{run} = 10\cdot t_{dyn}$, as well as  a second more luminous light curve  produced from a model assuming  a higher binary mass of 30 $M_{\odot}$.}
\label{Fig:bolocomp14ej}
\end{figure}

\clearpage
\input{tables/table1.tex}

\input{tables/table2.tex}

\clearpage
\input{tables/table3.tex}

\input{tables/table4.tex}

\clearpage 
\input{tables/table5.tex}

\input{tables/table6.tex}

\clearpage
\input{tables/table7.tex}

\input{tables/table8.tex}

\input{tables/table9.tex}

\clearpage
\input{tables/table10.tex}

\input{tables/table11.tex}

\clearpage
\begin{appendix}

\section{Metallicity and local environment}
\label{appendixA}

\begin{figure}[!htb]
\centering
\includegraphics[width=12cm]{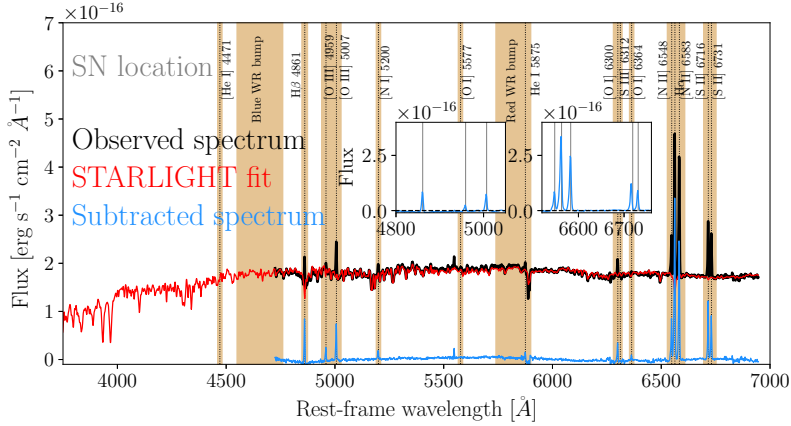}
\caption{ Visual wavelength MUSE spectrum at $+$390~d at the site of AT~2014ej in NGC~7552, shown in the host-galaxy rest frame and plotted in black.  Prominent host nebular emission features are shown in blue and are labeled, and each fit with a single-Gaussian function after the background computed by STARLIGHT (red line) was subtracted. The nebular lines that are also shown in the insets provide flux ratios  indicating a super-solar metallicity (12~$+$~log(O/H) $>$ 8.7$\pm$0.2 dex) on the O3N2 index at the location of AT~2014ej.} 
\label{fig:MUSEspectrum}
\end{figure}

To study the properties of the environment of AT~2014ej, we used integral field spectroscopic observations of NGC~7552 taken  with the European Southern Observatory's (ESO) Very Large Telescope (VLT) equipped with the instrument MUSE (Multi Unit Spectroscopic Explorer; \citealt{2014Msngr.157...13B}) on 2015 October  11 that are available from the ESO Science Archive.
The MUSE data set consist of a combination of three single pointings covering the same one squared arcmin field, with a total exposure of 3513 s. A one-dimensional extracted spectrum is plotted in Fig.~\ref{fig:MUSEspectrum}.

First, using a circular aperture of 2 arcsec diameter, we extracted a spectrum of the galaxy core and measured  the six strongest 
gas-phase emission line features in the spectra to obtain an average redshift of $z=0.00535\pm0.00003$, where the error is the mean error, which does not include the uncertainty related to the placement of the \ion{H}{ii} region in the galaxy.
This is fully consistent with the value listed in the NED.  

To estimate the local metallicity, we first extracted a circular spectrum with 2 arcsec diameter centered at the position of  AT~2014ej and performed a similar analysis as described by  \cite{2016MNRAS.455.4087G,2018ApJ...855..107G}. 
In summary, the spectrum was first fit with a modified version of {\sc STARLIGHT}.
The best-fit simple stellar populations (SSP) model was removed from the observed spectrum, which enabled us to obtain a pure gas-emission spectrum,  Next  the flux of the most prominent emission lines was estimated with Gaussian fits after correcting for dust attenuation. The adopted reddening value was derived from the Balmer decrement assuming case~B recombination \citep{2006agna.book.....O} and a   \cite{1999PASP..111...63F} extinction law characterized by an  $R_V=3.1$. 

From the extinction-corrected fluxes measured from H$\alpha$ $\lambda$6563, H$\beta$ $\lambda$4861, [\ion{O}{iii}] $\lambda$5007 , and the [\ion{N}{ii}] $\lambda$6583 emission lines, we  obtain a measure of the oxygen metallicity abundance on the  O3N2 index following the relation from \citet{2004MNRAS.348L..59P}. 
In doing so,  a metallicity higher than solar of 12~$+$~log(O/H) $= 8.7\pm$0.2 dex is inferred from the MUSE observations of NGC~7552 extracted at the position of AT~2014ej. 


\clearpage
\section{Galactic and host-galaxy reddening}
\label{appendixB}

\begin{figure}[!htb]
{\includegraphics[width=8cm]{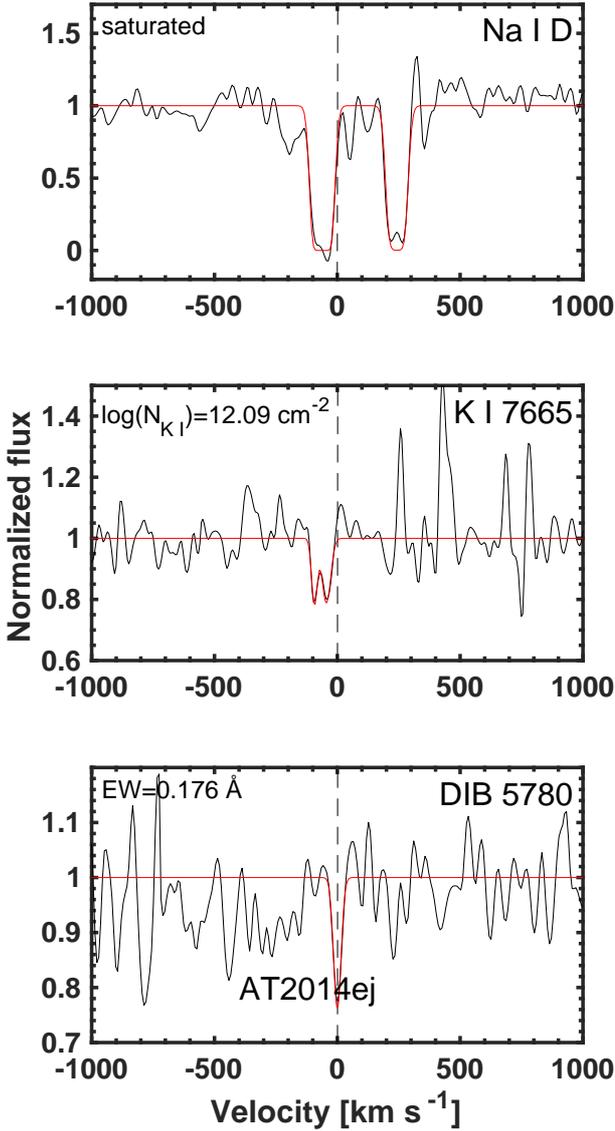}}
\caption{\texttt{VPFIT} models (red line) for the (top) $\ion{Na}{i}~D$, (middle) \ion{K}{i}, and (bottom)  DIB at 5780~\AA\ absorption features present in  the Magellan Clay ($+$ MIKE) medium-resolution spectrum (black line) of AT~2014ej discussed in detail in Appendix.~\ref{appendixB}.  According to \citet[][see their Fig.~3]{2013ApJ...779...38P}, the column density for  $\ion{Na}{i}~D$ is at the saturation limit and therefore does not provide a reliable estimate for reddening. The $\ion{K}{i}$ lines do provide an extinction estimate, but come with a high uncertainty. Finally, according to \citet[][see their Eq~6]{2013ApJ...779...38P}, the EW of the DIB feature at  5780~\AA\ estimated with \texttt{fitprofs} task in \texttt{IRAF} gives the visual extinction $A^{host}_{V} = 0.92\pm0.47$~mag. This value is fully consistent with the  extinction value estimated using the Balmer decrement combined with line measurements made from the MUSE spectrum (see Fig.~\ref{fig:MUSEspectrum}) at the location of AT~2014ej.}
\label{NGC7552-NaIDKI-fit}
\end{figure}

\newpage
The Milky Way reddening in the direction of AT~2014ej is very weak, with NED listing a \citet{2011ApJ...737..103S} color excess  $E(B-V)_{MW}=0.012$ mag, which, when a standard Galactic reddening law is adopted, corresponds to a visual extinction of $A_{V}^{MW} = 0.04$~mag. 
Inspection of low- and medium-resolution spectra presented in Appendix~\ref{appendixA} reveals conspicuous \ion{Na}{i}~D and also the \ion{K}{I} $\lambda\lambda$7665, 7699 interstellar lines located at the redshift of the host. Unfortunately, as demonstrated in Fig.~\ref{NGC7552-NaIDKI-fit} (top panel), the \ion{Na}{i}~D lines are saturated in our higher dispersion spectrum and therefore do not provide a reliable estimate of the host reddening.

The \ion{K}{I} lines in the MIKE spectrum that we show in Fig.~\ref{NGC7552-NaIDKI-fit} (middle panel) can be used to estimate the host extinction. 
To estimate the host extinction from these features, we made use of Eq.~(5) from \citet[][]{2013ApJ...779...38P}, which  connects the \ion{K}{i} column density to the host visual-extinction $A^{host}_{V}$. 
Column density estimates were obtained with the Voigt profile-fitting program \texttt{VPFIT}.\footnote{\url{http://www.ast.cam.ac.uk/rfc/vpfit.html}} 
The \texttt{VPFIT} fits to the data are shown Fig.~\ref{NGC7552-NaIDKI-fit} and provide column density values of log$_{10}(N_{\rm\ion{Na}{i}~D}$)$=12.09$~cm$^{-2}$.
Following Eq.~(5) of \citet[][]{2013ApJ...779...38P}, these  values correspond to $A^{host}_{V} = 2.5\pm1.8$~mag, where the accompanying uncertainties correspond to 72\% of the inferred value of $A^{host}_{V}$ \citep{2013ApJ...779...38P}.
This is a high extinction value with a  high  uncertainty. Therefore we resorted to  estimating the reddening based on a relation between the strength of the diffuse interstellar band (DIB) at 5780~\AA\ and $A_V$, again from  \citet{2013ApJ...779...38P}.

Inspection of our medium-dispersion  spectrum   reveals the 5780~\AA\ absorption feature (see Fig.~\ref{NGC7552-NaIDKI-fit}, bottom panel).
We computed the  equivalent width (EW) of the DIB feature by fitting a Gaussian function using the IRAF task \texttt{fitprofs}. 
In  computing the fit, an FWHM value was set to  2.1~\AA\  and the fitting range was set to be between  5778 -- 5782~\AA.
The Gaussian fit provides an EW of 0.176$\pm$0.020 \AA, which, when Eq.~(6) from \citet{2013ApJ...779...38P} is used, implies $A^{host}_{V} = 0.92\pm0.47$ mag. 
We verified that this value is fully consistent with the visual extinction of  $A^{host}_V=0.97\pm0.21$ mag implied by the Balmer decrement estimate from the MUSE spectrum presented in Appendix~\ref{appendixA} that we plot in Fig.~\ref{fig:MUSEspectrum}.
When we combine the DIB visual extinction estimate with the Milky Way contribution, we obtain a total visual extinction for AT~2014ej of $A^{tot}_{V} = 0.96\pm0.47$~mag, which is adopted in this work. Moreover, to convert $A^{tot}_{V}$ into extinction values appropriate for the different  passbands we used to obtain  photometry of AT~2014ej, we adopted a  \citet{NED_extinction} reddening law.
We finally note that  AT~2014ej  might   be associated with circumstellar dust,  and therefore the assumed host-reddening value comes with the same caveats as discussed in Appendix~B  of Paper~1 (see also \citealt{2012ApJ...759...20K}).

\end{appendix}

\end{document}

%% file: tables/table1.tex
\begin{deluxetable}{ccccccccc}
\tablewidth{0pt}
\tablecaption{Optical photometry of the local sequence for AT~2014ej in the `standard' system.\tablenotemark{a}\label{tab:CSPoptlocseq}}
\tablehead{
\colhead{ID} &
\colhead{$\alpha (2000)$} &
\colhead{$\delta (2000)$} &
\colhead{$B$} & 
\colhead{$V$} & 
\colhead{$u^{\prime}$} & 
\colhead{$g^{\prime}$} &
\colhead{$r^{\prime}$} & 
\colhead{$i^{\prime}$}}
\startdata
1 & 349.073883 & -42.532539 & $\ldots$     & $\ldots$     & $\ldots$     & $\ldots$     & $\ldots$     & $\ldots$ \\
2 & 349.146027 & -42.515438 & $\cdots$     & $\ldots$     & $\ldots$     & $\ldots$     & $\ldots$     & $\ldots$ \\
3 & 348.886597 & -42.615513 & 14.807(010)& 14.145(007)& 15.940(034)& 14.434(007)& 13.949(005)& 13.805(007)\\
  4 & 349.104828 & -42.476219 & 15.332(013)& 14.847(011)& 16.056(036)& 15.053(009)& 14.713(007)& 14.605(011)\\
  5 & 348.940552 & -42.641197 & 15.983(037)& 15.050(030)& 17.578(091)& 15.504(031)& 14.689(029)& 14.316(014)\\
  6 & 349.160797 & -42.579163 & 15.729(022)& 15.144(010)& 16.608(051)& 15.402(011)& 14.974(008)& 14.828(007)\\
  7 & 349.018341 & -42.487396 & 15.825(007)& 15.114(011)& 16.957(050)& 15.434(012)& 14.883(008)& 14.683(010)\\
  8 & 349.177582 & -42.494061 & 16.206(039)& 15.209(009)& 18.097(048)& 15.689(015)& 14.782(010)& 14.447(009)\\
  9 & 349.201538 & -42.602962 & 15.843(024)& 15.269(010)& 16.636(050)& 15.515(009)& 15.114(009)& 14.967(011)\\
 10 & 348.887085 & -42.535305 & 16.070(021)& 15.241(010)& 17.454(089)& 15.626(013)& 14.936(008)& 14.675(008)\\
 11 & 349.011200 & -42.468658 & 15.985(017)& 15.422(008)& 16.833(051)& 15.671(012)& 15.273(011)& 15.141(012)\\
 12 & 348.950928 & -42.493233 & 16.349(015)& 15.528(011)& 17.800(063)& 15.903(011)& 15.248(013)& 15.027(011)\\
 13 & 349.079193 & -42.577908 & 16.528(029)& 15.614(012)& 17.847(229)& 16.049(013)& 15.289(008)& 14.948(010)\\
 14 & 348.948059 & -42.539196 & 16.196(030)& 15.621(013)& 17.028(082)& 15.875(015)& 15.465(016)& 15.316(011)\\
 15 & 349.177124 & -42.531296 & 16.463(027)& 15.645(015)& 17.914(059)& 16.018(015)& 15.356(012)& 15.141(009)\\
 16 & 349.107941 & -42.547867 & 16.849(022)& 16.196(028)& 17.786(073)& 16.488(034)& 15.999(016)& 15.842(022)\\
 17 & 349.041046 & -42.637428 & 18.129(070)& 17.004(051)& 19.990(061)& 17.570(043)& 16.477(030)& 16.051(021)\\
 18 & 349.142731 & -42.624664 & 17.243(031)& 16.379(035)& 18.671(075)& 16.785(021)& 16.065(016)& 15.824(018)\\
 19 & 349.086792 & -42.654221 & 17.109(047)& 16.521(032)& 17.911(083)& 16.773(014)& 16.358(019)& 16.216(036)\\
 20 & 349.004547 & -42.548927 & 17.406(080)& 16.513(033)& 18.663(085)& 16.931(037)& 16.227(014)& 15.914(024)\\
 21 & 349.021851 & -42.545380 & 17.152(040)& 16.558(030)& 17.871(092)& 16.843(046)& 16.377(019)& 16.174(017)\\
 22 & 349.063202 & -42.463596 & 17.234(067)& 16.593(036)& 18.179(051)& 16.868(033)& 16.405(032)& 16.243(030)\\
 23 & 349.155212 & -42.636612 & 17.182(050)& 16.607(023)& 17.946(056)& 16.847(031)& 16.477(013)& 16.358(019)\\
 24 & 348.945343 & -42.460548 & 17.043(035)& 16.618(040)& 17.523(170)& 16.811(024)& 16.502(037)& 16.415(061)\\
 25 & 349.102997 & -42.480492 & 17.794(058)& 16.606(041)& 19.879(052)& 17.199(042)& 16.088(015)& 15.623(015)\\
 26 & 349.023193 & -42.670746 & 17.517(080)& 16.726(046)& 18.875(074)& 17.077(049)& 16.433(016)& 16.205(022)\\
 27 & 349.009583 & -42.517628 & 17.768(078)& 16.937(036)& 19.021(070)& 17.327(021)& 16.621(039)& 16.345(027)\\
 28 & 348.988068 & -42.467167 & 17.582(100)& 16.962(035)& 18.510(044)& 17.236(044)& 16.762(023)& 16.626(037)\\
\enddata
\tablenotetext{a}{Note. -- Values in parenthesis are 1-$\sigma$ uncertainties  that correspond to the rms of the instrumental errors of the photometry obtained over a minimum of three nights observed relative to standard star fields.}
\end{deluxetable}

%% file: tables/table2.tex
\begin{deluxetable}{ccccccccc}
\tablecolumns{9}
\tablewidth{0pt}
\tablecaption{NIR photometry of the local sequences for AT~2014ej in the `standard' system.\tablenotemark{a}\label{tab:AT2014ejnirlocseq}}
\tablehead{
\colhead{ID} &
\colhead{$\alpha (2000)$} &
\colhead{$\delta (2000)$} &
\colhead{$Y$} &
\colhead{N}   & 
\colhead{$J_{rc2}$} &
\colhead{N}   & 
\colhead{$H$} &
\colhead{N}   }

\startdata
101 & 349.044870 & -42.571004 & 15.64(02) & 3 & 15.38(04) & 3 & 15.07(07) &	3\\ 
102 & 349.018801 & -42.593902 & 16.57(06) &	3 & 16.11(06) & 3 & 15.61(16) &	3\\ 
103 & 349.015365 & -42.595754 & 17.60(07) &	2 & $\cdots$    & $\cdots$  & $\cdots$   &	$\cdots$\\ 
104 & 349.061355 & -42.592496 & 17.52(41) &	2 & 17.08(21) & 3 & 17.01(15) &	1\\ 
106 & 349.068255 & -42.572860 & 18.37(11) &	1 & $\cdots$    & $\cdots$ & $\cdots $   &	$\cdots$\\ 
107 & 349.059690 & -42.570588 & 18.29(10) &	1 & $\cdots$    & $\cdots$ & $\cdots $   &	$\cdots$\\ 
109 & 349.006081 & -42.561110 & 18.54(12) &	1 & $\cdots$    & $\cdots$ & $\cdots $   &	$\cdots$\\  
\enddata
\tablenotetext{a}{
Note. -- Uncertainties given in parentheses in thousandths of a magnitude correspond to an rms of the magnitudes obtained on photometric nights.
}
\end{deluxetable}

%% file: tables/table3.tex
\begin{deluxetable}{cc|cc|cc|cc|cc|cc}
\rotate
\tabletypesize{\scriptsize}
\tablewidth{0pt}
\tablecolumns{18}
\tablecaption{Optical photometry of AT~2014ej in the Swope `natural' system.\label{CSPII-optphot}}
\tablehead{
\colhead{Time\tablenotemark{a}}&
\colhead{$u$}&
\colhead{Time\tablenotemark{a}}&
\colhead{$B$}&
\colhead{Time\tablenotemark{a}}&
\colhead{$g$}&
\colhead{Time\tablenotemark{a}}&
\colhead{$V$}&
\colhead{Time\tablenotemark{a}}&
\colhead{$r$}&
\colhead{Time\tablenotemark{a}}&
\colhead{$i$}}
\startdata
     $\cdots$       &$\cdots$ & 25.71 & 18.472(0.038)  & 25.70 & 18.178(0.033) & 25.71 & 18.022(0.035)  & 25.70 & 17.739(0.035) &  25.71 & 17.702(0.037) \\
	 26.64 & 19.226(0.040) & 26.62 & 18.556(0.035)  & 26.63 & 18.297(0.034) & 26.63 & 18.085(0.037) & 26.63 & 17.829(0.031) &  26.63 & 17.801(0.029) \\
     27.62 & 19.322(0.036) & 27.63 & 18.619(0.037)  & 27.61 & 18.402(0.045) & 27.63 & 18.201(0.037) & 27.61 & 17.962(0.037) &  27.62 & 17.985(0.034) \\
     28.62 & 19.468(0.050) & 28.61 & 18.737(0.036)  & 28.61 & 18.542(0.035) & 28.61 & 18.266(0.043) & 28.61 & 18.023(0.029) &  28.61 & 17.953(0.032) \\
     29.61 & 19.509(0.048) & 29.62 & 18.835(0.032)  & 29.59 & 18.544(0.061) & 29.62 & 18.372(0.044) & 29.60 & 18.144(0.037) &  29.60 & 18.073(0.024) \\ 
     30.59 & 19.586(0.034) & 30.61 & 18.900(0.033)  & 30.60 & 18.604(0.032) & 30.61 & 18.387(0.033) & 30.59 & 18.190(0.034) &  30.60 & 18.088(0.035) \\
     31.64 & 19.753(0.052) & 31.63 & 19.012(0.043)  & 31.64 & 18.671(0.037) & 31.63 & 18.504(0.044) & 31.64 & 18.210(0.041) &  31.64 & 18.180(0.039) \\ 
     32.63 & 19.920(0.064) & 32.62 & 19.016(0.034)  & 32.63 & 18.729(0.039) & 32.62 & 18.524(0.033) & 32.63 & 18.351(0.033) &  32.64 & 18.219(0.040) \\ 
     36.60 & 20.201(0.125) & 36.59 & 19.294(0.060)  & 36.60 & 18.918(0.040) & 36.59 & 18.796(0.046) & 36.60 & 18.548(0.037) &  36.60 & 18.394(0.043) \\
     37.63 & 20.232(0.101) & 37.64 & 19.347(0.059)  & $\cdots$  &  $\cdots$ & 37.63 & 18.734(0.042) & 37.62 & 18.515(0.038) &  37.62 & 18.468(0.056)\\
     38.65 & 20.038(0.115) & 38.63 & 19.399(0.069)  & 38.64 & 18.969(0.042) & 38.63 & 18.726(0.037) & 38.64 & 18.487(0.033) &  38.64 & 18.385(0.040) \\
     39.65 & 20.276(0.109) & 39.63 & 19.332(0.048)  & 39.64 & 19.015(0.043) & 39.64 & 18.688(0.047) & 39.64 & 18.456(0.034) &  39.64 & 18.380(0.044) \\
     40.62 & 20.427(0.107) & 40.60 & 19.459(0.036)  & 40.61 & 18.967(0.029) & 40.61 & 18.803(0.042) & 40.61 & 18.484(0.035) &  40.61 & 18.381(0.048) \\
     41.61 & 20.562(0.166) & 41.62 & 19.439(0.039)  & 41.60 & 19.096(0.042) & 41.62 & 18.810(0.042) & 41.60 & 18.607(0.036) &  41.60 & 18.376(0.038) \\
           \enddata
\tablenotetext{a}{Note. -- JD+2456900.}
\end{deluxetable}

%% file: tables/table4.tex
\begin{deluxetable}{c c c c}
	\tablecaption{NIR photometry of AT~2014ej in the du Pont (+ RetroCam) `natural' system.\tablenotemark{a} \label{tab:AT2014ej_nirphot}}
	\tablecolumns{4}
	\tablewidth{0pt}
	\tablehead{
\colhead{JD}  & 
\colhead{photometry} & 
\colhead{error} &
\colhead{filter }}
\startdata
2456929.64 & 17.51 & 0.02 &  $    Y$ \\  
2456931.58 & 17.58 & 0.03 &  $    Y$ \\
2456932.58 & 17.68 & 0.03 &  $    Y$ \\
2456929.67 & 17.24 & 0.02 &  $ J_{rc2}$ \\
2456931.62 & 17.35 & 0.03 &  $ J_{rc2}$ \\
2456932.61 & 17.47 & 0.03 &  $ J_{rc2}$ \\
2456929.66 & 16.79 & 0.03 &  $    H$ \\
2456931.61 & 16.92 & 0.03 &  $    H$ \\
2456932.60 & 17.05 & 0.03 &  $    H$ \\
2456937.59 & 16.17 & 0.14 &  $K_s$\tablenotemark{b}  \\ \enddata
\tablenotetext{a}{Note. -- The uncertainties in photometry correspond to the sum in quadrature of the instrumental photometric error and the nightly zero-point error.}
\tablenotetext{b}{Note. -- Observations were obtained with the BAADE (+ FourStar) telescope, and the photometric zero-point was computed relative to two 2MASS stars. Also note that no template subtraction was performed on this single science image.}
	\end{deluxetable}

%% file: tables/table5.tex
\begin{deluxetable}{ccc}
\tablewidth{0pt}
\tablecaption{CHASE unfiltered filter photometry of AT~2014ej.\label{tab:AT2014ej_CHASE_phot}}
\tablehead{
\colhead{MJD} &
\colhead{$O$\tablenotemark{a}}  &
\colhead{Telescope}}
\startdata
56910.27  &    $>$18.3     &PROMPT4\\
56920.06  &    17.519(054) &PROMPT1\\  
56930.32  &    18.286(121) &PROMPT4\\
56940.05  &    18.668(165) &PROMPT1\\
56941.27  &    18.656(318) &PROMPT3\\
56943.04  &    18.699(158) &PROMPT4\\
56943.27  &    18.527(090) &PROMPT1\\
56956.24  &    18.217(118) &PROMPT4\\
56958.10  &    18.043(072) &PROMPT1\\
56959.23  &    18.106(088) &PROMPT4\\
56961.05  &    18.119(085) &PROMPT4\\
56978.13  &    17.669(102) &PROMPT4\\
56983.07  &    17.886(079) &PROMPT4\\
56998.12  &    18.025(148) &PROMPT4\\
57015.08  &    18.361(124) &PROMPT3\\
\enddata
\tablenotetext{a}{Note. -- Unfiltered photometry calibrated relative to the CSP-II $r$-band local sequence.}
\end{deluxetable}

%% file: tables/table6.tex
\begin{deluxetable}{c c c c}
	\tablecaption{NIR photometry of AT~2014ej from PESSTO.\tablenotemark{a} \label{tab:AT2014ej_nirphot_pessto}}
	\tablecolumns{4}
	\tablewidth{0pt}
	\tablehead{
\colhead{MJD}  & 
\colhead{photometry} & 
\colhead{error} &
\colhead{filter}}
\startdata
56944.163 & 17.07 &0.08 &  $J$\\
56944.163 & 16.97 &0.09 &  $H$\\
56944.163 & 16.37 &0.08 &  $K$\\
56954.000 & 16.39 &0.05 &  $K$\\
56964.250 & 16.09 &0.08 &  $K$\\
\enddata
\tablenotetext{a}{Note. -- Observations were obtained with the NTT(+SOFI) telescope, and the photometric zero-point was computed relative to 2MASS stars.} 
\end{deluxetable}

%% file: tables/table7.tex
\clearpage 
\begin{deluxetable}{l l l c l l c}
\tablewidth{0pt}
\tablecolumns{7}
\tablecaption{\label{tab:specobs} Journal of Spectroscopic Observations of AT~2014ej.}
\tablehead{
\colhead{Object} & 
\colhead{Date} & 
\colhead{Date} &	
\colhead{Days since} & 
\colhead{Telescope} & 
\colhead{Instrument}  & 
\colhead{Resolution}\\
\colhead{} & 
\colhead{(JD)} &
\colhead{(UT)} & 
\colhead{discovery\tablenotemark{a}} &
\colhead{} &
\colhead{}   & 
\colhead{(\AA)}}
\startdata
AT~2014ej & 2456927.67 & September 27.16 & $+$7.1 & du Pont& WFCCD &   7.5\\
AT~2014ej & 2456930.60 & September 30.10 & $+$10.0 & NTT    & EFOSC2 &  15.8/16.0\tablenotemark{b}\\
AT~2014ej & 2456931.59 & October 01.09   & $+$10.6 & Clay   & MIKE  &  0.21\\
AT~2014ej & 2456933.96 & October 03.46   & $+$13.3 & ANU   & WiFeS  &  0.98\\
AT~2014ej & 2456953.62 & October 23.12   & $+$32.9 & NTT    & EFOSC2 &  15.8/16.0 \\ 
AT~2014ej & 2456963.61 & November 02.11  & $+$42.8 & NTT    & EFOSC2 &  15.8/16.0 \\ 
AT~2014ej & 2456992.57 & December 01.07  & $+$71.6 & NTT    & EFOSC2 & 15.8/16.0\ \\ 
\enddata
\tablenotetext{a}{Days since outburst assuming outburst date for AT~2014ej of 24.06 September 2014 UT (JD$-$2456924.56).}
\tablenotetext{b}{Note. -- NTT (+ EFOSC) observations were obtained with two grisms, i.e., Gr{\#}11 and Gr{\#}16, providing slightly different spectral resolutions.}
\end{deluxetable}

%% file: tables/table8.tex
\begin{deluxetable}{ccc}
\tablewidth{0pt}
\tablecaption{PESSTO $V$-band photometry from acquisition images of AT~2014ej.\label{tab:AT2014ej_pessto_phot}}
\tablehead{
\colhead{MJD} &
\colhead{$V$\tablenotemark{a}}  &
\colhead{Telescope}}
\startdata
56930.095 & 18.458(0.031) & NTT\\
56953.095 & 18.566(0.036) & NTT\\
56963.080 & 18.389(0.037) & NTT\\
56992.070 & 18.195(0.045) & NTT\\
\enddata
\tablenotetext{a}{Note. -- $V$-band photometry calibrated relative to the CSP-II $V$-band local sequence listed in Table~\ref{tab:CSPoptlocseq}.}
\end{deluxetable}

%% file: tables/table9.tex
\begin{deluxetable}{c|c|c|c|c|}
\rotate
\tabletypesize{\scriptsize}
\tablewidth{0pt}

\tablecaption{Spectrophotometry of AT~2014ej in the Swope `natural' system.\label{tab:spectrophotometry}}
\tablehead{
\colhead{JD}&
\colhead{$B$}&
\colhead{$g$}&
\colhead{$r$}&
\colhead{$i$}}
\startdata
2456927.67 &  18.709   &    18.446    &   17.967    &  17.898 \\
2456930.60 &  18.821   &    18.574    &   18.156    &  \ldots \\
2456933.96 &  \ldots   &    \ldots    &   18.449    &  18.425 \\
2456953.62 &  19.199   &    18.904    &   18.324    &  18.209 \\
2456963.61 &  18.992   &    18.693    &   18.166    &  18.005 \\
2456992.57 &  18.776   &    18.496    &   17.959    &  17.802 \\
\enddata
\tablecomments{We assume a conservative error of 0.1 mag on the spectrophotometry due
to the fact that there might be host galaxy flux contaminating the spectra.}
\end{deluxetable}

%% file: tables/table10.tex
\begin{deluxetable}{lcc}
\tablewidth{0pt}
\tablecaption{\label{tab:comp_obj_at2014ej} Adopted color excess values and distance moduli for the LRN sample compared to AT~2014ej.}
\tablehead{
\colhead{LRN}& 
\colhead{$\mu$ } & 
\colhead{$E(B-V)_{tot}$}  \\
\colhead{} & 
\colhead{(mag)} & 
\colhead{(mag)} }
\startdata
 SN~1997bs           & 29.82  & 0.21    \\
 NGC~3437-2011-OT1   & 31.60  & 0.02    \\
 NGC~4490-2011-OT1   & 29.91  & 0.32    \\
 UGC~12307-2013-OT1  & 32.99  & 0.22    \\
 M101-2015-OT1       & 29.04  & 0.008   \\
 SNhunt248           & 31.76  & 0.045   \\
 AT~2017jfs          & 32.73  & 0.022   \\
 AT~2018hso          & 31.64  & 0.30    \\
\enddata
\end{deluxetable}

%% file: tables/table11.tex
\begin{deluxetable}{lcc}
\tablecolumns{9}
\tablewidth{0pt}
\tablecaption{\label{tab:FWHM_AT2014ej} FWHM velocity measurements of prominent spectral features in AT~2014ej.}
\tablehead{
\colhead{Phase\tablenotemark{a}} &
\colhead{H$\beta$} &
\colhead{H$\alpha$} \\
\colhead{(d) } &
\colhead{(km s$^{-1}$)} &
\colhead{(km s$^{-1}$) }}
\startdata
+7   & 854$\pm$72  & 107$\pm$18\\
+10  & 868$\pm$114  & 969$\pm$78\\
+33 & $\cdots$     & 999$\pm$75\\
+43 & $\cdots$     & 843$\pm$70\\
+72 & $\cdots$     & 1115$\pm$117\\
\enddata
\tablenotetext{a}{Note. -- Phase with respect to date of discovery.}
\end{deluxetable}

%% file: 38019corr_max.bbl
\begin{thebibliography}{76}
\expandafter\ifx\csname natexlab\endcsname\relax\def\natexlab#1{#1}\fi

\bibitem[{{Adams} {et~al.}(2016){Adams}, {Kochanek}, {Prieto}, {Dai},
  {Shappee}, \& {Stanek}}]{2016MNRAS.460.1645A}
{Adams}, S.~M., {Kochanek}, C.~S., {Prieto}, J.~L., {et~al.} 2016, \mnras, 460,
  1645

\bibitem[{{Bacon} {et~al.}(2014){Bacon}, {Vernet}, {Borisova}, {Bouch{\'e}},
  {Brinchmann}, {Carollo}, {Carton}, {Caruana}, {Cerda}, {Contini}, {Franx},
  {Girard}, {Guerou}, {Haddad}, {Hau}, {Herenz}, {Herrera}, {Husemann},
  {Husser}, {Jarno}, {Kamann}, {Krajnovic}, {Lilly}, {Mainieri}, {Martinsson},
  {Palsa}, {Patricio}, {P{\'e}contal}, {Pello}, {Piqueras}, {Richard},
  {Sandin}, {Schroetter}, {Selman}, {Shirazi}, {Smette}, {Soto}, {Streicher},
  {Urrutia}, {Weilbacher}, {Wisotzki}, \& {Zins}}]{2014Msngr.157...13B}
{Bacon}, R., {Vernet}, J., {Borisova}, E., {et~al.} 2014, The Messenger, 157,
  13

\bibitem[{{Bernstein} {et~al.}(2003){Bernstein}, {Shectman}, {Gunnels},
  {Mochnacki}, \& {Athey}}]{2003SPIE.4841.1694B}
{Bernstein}, R., {Shectman}, S.~A., {Gunnels}, S.~M., {Mochnacki}, S., \&
  {Athey}, A.~E. 2003, in \procspie, Vol. 4841, Instrument Design and
  Performance for Optical/Infrared Ground-based Telescopes, ed. M.~{Iye} \&
  A.~F.~M. {Moorwood}, 1694--1704

\bibitem[{{Blagorodnova} {et~al.}(2020){Blagorodnova}, {Karambelkar}, {Adams},
  {Kasliwal}, {Kochanek}, {Dong}, {Campbell}, {Hodgkin}, {Jencson},
  {Johansson}, {Kozlowski}, {Laher}, {Masci}, {Nugent}, \&
  {Rebbapragada}}]{Blagorodnova2020}
{Blagorodnova}, N., {Karambelkar}, V., {Adams}, S.~M., {et~al.} 2020, arXiv
  e-prints, arXiv:2004.04757

\bibitem[{{Blagorodnova} {et~al.}(2017){Blagorodnova}, {Kotak}, {Polshaw},
  {Kasliwal}, {Cao}, {Cody}, {Doran}, {Elias-Rosa}, {Fraser}, {Fremling},
  {Gonzalez-Fernandez}, {Harmanen}, {Jencson}, {Kankare}, {Kudritzki},
  {Kulkarni}, {Magnier}, {Manulis}, {Masci}, {Mattila}, {Nugent}, {Ochner},
  {Pastorello}, {Reynolds}, {Smith}, {Sollerman}, {Taddia}, {Terreran},
  {Tomasella}, {Turatto}, {Vreeswijk}, {Wozniak}, \&
  {Zaggia}}]{2017ApJ...834..107B}
{Blagorodnova}, N., {Kotak}, R., {Polshaw}, J., {et~al.} 2017, \apj, 834, 107

\bibitem[{{Bock} {et~al.}(2014){Bock}, {Marples}, {Parker}, {Morrell},
  {Contreras}, {Gonzalez}, {Phillips}, {Stritzinger}, {Hsiao}, {Gall},
  {Holmbo}, {Taddia}, {Sanchez}, \& {Lira}}]{2014CBET.3998....1B}
{Bock}, G., {Marples}, P., {Parker}, S., {et~al.} 2014, Central Bureau
  Electronic Telegrams, 3998, 1

\bibitem[{{Botticella} {et~al.}(2009){Botticella}, {Pastorello}, {Smartt},
  {Meikle}, {Benetti}, {Kotak}, {Cappellaro}, {Crockett}, {Mattila}, {Sereno},
  {Patat}, {Tsvetkov}, {van Loon}, {Abraham}, {Agnoletto}, {Arbour}, {Benn},
  {di Rico}, {Elias-Rosa}, {Gorshanov}, {Harutyunyan}, {Hunter}, {Lorenzi},
  {Keenan}, {Maguire}, {Mendez}, {Mobberley}, {Navasardyan}, {Ries},
  {Stanishev}, {Taubenberger}, {Trundle}, {Turatto}, \&
  {Volkov}}]{2009MNRAS.398.1041B}
{Botticella}, M.~T., {Pastorello}, A., {Smartt}, S.~J., {et~al.} 2009, \mnras,
  398, 1041

\bibitem[{{Burns} {et~al.}(2018){Burns}, {Parent}, {Phillips}, {Stritzinger},
  {Krisciunas}, {Suntzeff}, {Hsiao}, {Contreras}, {Anais}, {Boldt}, {Busta},
  {Campillay}, {Castell{\'o}n}, {Folatelli}, {Freedman}, {Gonz{\'a}lez},
  {Hamuy}, {Heoflich}, {Krzeminski}, {Madore}, {Morrell}, {Persson}, {Roth},
  {Salgado}, {Ser{\'o}n}, \& {Torres}}]{2018ApJ...869...56B}
{Burns}, C.~R., {Parent}, E., {Phillips}, M.~M., {et~al.} 2018, \apj, 869, 56

\bibitem[{{Cai} {et~al.}(2019){Cai}, {Pastorello}, {Fraser}, {Prentice},
  {Reynolds}, {Cappellaro}, {Benetti}, {Morales-Garoffolo}, {Reguitti},
  {Elias-Rosa}, {Brennan}, {Callis}, {Cannizzaro}, {Fiore}, {Gromadzki},
  {Galindo-Guil}, {Gall}, {Heikkil{\"a}}, {Mason}, {Moran}, {Onori},
  {Sagu{\'e}s Carracedo}, \& {Valerin}}]{at2018hso}
{Cai}, Y.-Z., {Pastorello}, A., {Fraser}, M., {et~al.} 2019, arXiv e-prints,
  arXiv:1909.13147

\bibitem[{{Childress} {et~al.}(2016){Childress}, {Tucker}, {Yuan}, {Scalzo},
  {Ruiter}, {Seitenzahl}, {Zhang}, {Schmidt}, {Anguiano}, {Aniyan}, {Bayliss},
  {Bento}, {Bessell}, {Bian}, {Davies}, {Dopita}, {Fogarty}, {Fraser-McKelvie},
  {Freeman}, {Kuruwita}, {Medling}, {Murphy}, {Murphy}, {Owers}, {Panther},
  {Sweet}, {Thomas}, \& {Zhou}}]{childress2016}
{Childress}, M.~J., {Tucker}, B.~E., {Yuan}, F., {et~al.} 2016, \pasa, 33, e055

\bibitem[{{Chugai} \& {Danziger}(1994)}]{1994MNRAS.268..173C}
{Chugai}, N.~N. \& {Danziger}, I.~J. 1994, \mnras, 268, 173

\bibitem[{{Dahlem}(2005)}]{grus_quartet}
{Dahlem}, M. 2005, \aap, 429, L5

\bibitem[{{Doherty} {et~al.}(2017){Doherty}, {Gil-Pons}, {Siess}, \&
  {Lattanzio}}]{2017PASA...34...56D}
{Doherty}, C.~L., {Gil-Pons}, P., {Siess}, L., \& {Lattanzio}, J.~C. 2017,
  \pasa, 34, e056

\bibitem[{{Filippenko}(1997)}]{1997ARA&A..35..309F}
{Filippenko}, A.~V. 1997, \araa, 35, 309

\bibitem[{{Fitzpatrick}(1999)}]{1999PASP..111...63F}
{Fitzpatrick}, E.~L. 1999, \pasp, 111, 63

\bibitem[{Fitzpatrick(1999)}]{NED_extinction}
Fitzpatrick, E.~L. 1999, PASP, 111, 63

\bibitem[{{Galbany} {et~al.}(2016){Galbany}, {Anderson}, {Rosales-Ortega},
  {Kuncarayakti}, {Kr{\"u}hler}, {S{\'a}nchez}, {Falc{\'o}n-Barroso},
  {P{\'e}rez}, {Maureira}, {Hamuy}, {Gonz{\'a}lez-Gait{\'a}n}, {F{\"o}rster},
  \& {Moral}}]{2016MNRAS.455.4087G}
{Galbany}, L., {Anderson}, J.~P., {Rosales-Ortega}, F.~F., {et~al.} 2016,
  \mnras, 455, 4087

\bibitem[{{Galbany} {et~al.}(2018){Galbany}, {Anderson}, {S{\'a}nchez},
  {Kuncarayakti}, {Pedraz}, {Gonz{\'a}lez-Gait{\'a}n}, {Stanishev},
  {Dom{\'\i}nguez}, {Moreno-Raya}, {Wood-Vasey}, {Mour{\~a}o}, {Ponder},
  {Badenes}, {Moll{\'a}}, {L{\'o}pez-S{\'a}nchez}, {Rosales-Ortega},
  {V{\'\i}lchez}, {Garc{\'\i}a-Benito}, \& {Marino}}]{2018ApJ...855..107G}
{Galbany}, L., {Anderson}, J.~P., {S{\'a}nchez}, S.~F., {et~al.} 2018, \apj,
  855, 107

\bibitem[{{Glebbeek} {et~al.}(2013){Glebbeek}, {Gaburov}, {Portegies Zwart}, \&
  {Pols}}]{ggpp13}
{Glebbeek}, E., {Gaburov}, E., {Portegies Zwart}, S., \& {Pols}, O.~R. 2013,
  \mnras, 434, 3497

\bibitem[{{Hashimoto} {et~al.}(1993){Hashimoto}, {Iwamoto}, \&
  {Nomoto}}]{1993ApJ...414L.105H}
{Hashimoto}, M., {Iwamoto}, K., \& {Nomoto}, K. 1993, \apjl, 414, L105

\bibitem[{{Howitt} {et~al.}(2019){Howitt}, {Stevenson}, {Vigna-G{\'o}mez},
  {Justham}, {Ivanova}, {Woods}, {Neijssel}, \& {Mandel}}]{hsv+19}
{Howitt}, G., {Stevenson}, S., {Vigna-G{\'o}mez}, A.~r., {et~al.} 2019, arXiv
  e-prints, arXiv:1912.07771

\bibitem[{{Humphreys} {et~al.}(2011){Humphreys}, {Bond}, {Bedin}, {Bonanos},
  {Davidson}, {Berto Monard}, {Prieto}, \& {Walter}}]{2011ApJ...743..118H}
{Humphreys}, R.~M., {Bond}, H.~E., {Bedin}, L.~R., {et~al.} 2011, \apj, 743,
  118

\bibitem[{{Ivanova} {et~al.}(2013){Ivanova}, {Justham}, {Chen}, {De Marco},
  {Fryer}, {Gaburov}, {Ge}, {Glebbeek}, {Han}, {Li}, {Lu}, {Marsh},
  {Podsiadlowski}, {Potter}, {Soker}, {Taam}, {Tauris}, {van den Heuvel}, \&
  {Webbink}}]{ijc+13}
{Ivanova}, N., {Justham}, S., {Chen}, X., {et~al.} 2013, \aapr, 21, 59

\bibitem[{{Kankare} {et~al.}(2015){Kankare}, {Kotak}, {Pastorello}, {Fraser},
  {Mattila}, {Smartt}, {Bruce}, {Chambers}, {Elias-Rosa}, {Flewelling},
  {Fremling}, {Harmanen}, {Huber}, {Jerkstrand}, {Kangas}, {Kuncarayakti},
  {Magee}, {Magnier}, {Polshaw}, {Smith}, {Sollerman}, \&
  {Tomasella}}]{kankare_snhunt248}
{Kankare}, E., {Kotak}, R., {Pastorello}, A., {et~al.} 2015, \aap, 581, L4

\bibitem[{{Kashi} {et~al.}(2010){Kashi}, {Frankowski}, \& {Soker}}]{kashi2010}
{Kashi}, A., {Frankowski}, A., \& {Soker}, N. 2010, \apjl, 709, L11

\bibitem[{{Kashi} \& {Soker}(2016)}]{kashi2016}
{Kashi}, A. \& {Soker}, N. 2016, Research in Astronomy and Astrophysics, 16, 99

\bibitem[{{Kiewe} {et~al.}(2012){Kiewe}, {Gal-Yam}, {Arcavi}, {Leonard},
  {Emilio Enriquez}, {Cenko}, {Fox}, {Moon}, {Sand }, {Soderberg}, \&
  {CCCP}}]{kiewe12IIn}
{Kiewe}, M., {Gal-Yam}, A., {Arcavi}, I., {et~al.} 2012, \apj, 744, 10

\bibitem[{{Kitaura} {et~al.}(2006){Kitaura}, {Janka}, \&
  {Hillebrandt}}]{2006A&A...450..345K}
{Kitaura}, F.~S., {Janka}, H.-T., \& {Hillebrandt}, W. 2006, \aap, 450, 345

\bibitem[{{Kochanek}(2011)}]{2011ApJ...741...37K}
{Kochanek}, C.~S. 2011, \apj, 741, 37

\bibitem[{{Kochanek} {et~al.}(2012{\natexlab{a}}){Kochanek}, {Khan}, \&
  {Dai}}]{2012ApJ...759...20K}
{Kochanek}, C.~S., {Khan}, R., \& {Dai}, X. 2012{\natexlab{a}}, \apj, 759, 20

\bibitem[{{Kochanek} {et~al.}(2012{\natexlab{b}}){Kochanek}, {Szczygie{\l}}, \&
  {Stanek}}]{2012ApJ...758..142K}
{Kochanek}, C.~S., {Szczygie{\l}}, D.~M., \& {Stanek}, K.~Z.
  2012{\natexlab{b}}, \apj, 758, 142

\bibitem[{{Kulkarni} \& {Kasliwal}(2009)}]{2009aaxo.conf..312K}
{Kulkarni}, S. \& {Kasliwal}, M.~M. 2009, in Astrophysics with All-Sky X-Ray
  Observations, ed. N.~{Kawai}, T.~{Mihara}, M.~{Kohama}, \& M.~{Suzuki}, 312

\bibitem[{{Lipunov} {et~al.}(2017){Lipunov}, {Blinnikov}, {Gorbovskoy},
  {Tutukov}, {Baklanov}, {Krushinski}, {Tiurina}, {Balanutsa}, {Kuznetsov},
  {Kornilov}, {Gorbunov}, {Shumkov}, {Vladimirov}, {Gress}, {Budnev}, {Ivanov},
  {Tlatov}, {Gabovich}, {Yurkov}, {Sergienko}, \&
  {Zalozhnykh}}]{2017MNRAS.470.2339L}
{Lipunov}, V.~M., {Blinnikov}, S., {Gorbovskoy}, E., {et~al.} 2017, \mnras,
  470, 2339

\bibitem[{{MacLeod} {et~al.}(2017){MacLeod}, {Macias}, {Ramirez-Ruiz},
  {Grindlay}, {Batta}, \& {Montes}}]{mmr+17}
{MacLeod}, M., {Macias}, P., {Ramirez-Ruiz}, E., {et~al.} 2017, \apj, 835, 282

\bibitem[{{MacLeod} {et~al.}(2018){MacLeod}, {Ostriker}, \& {Stone}}]{mos18}
{MacLeod}, M., {Ostriker}, E.~C., \& {Stone}, J.~M. 2018, \apj, 863, 5

\bibitem[{{Mauerhan} {et~al.}(2018){Mauerhan}, {Van Dyk}, {Johansson}, {Fox},
  {Filippenko}, \& {Graham}}]{2018MNRAS.473.3765M}
{Mauerhan}, J.~C., {Van Dyk}, S.~D., {Johansson}, J., {et~al.} 2018, \mnras,
  473, 3765

\bibitem[{{Metzger} \& {Pejcha}(2017)}]{2017MNRAS.471.3200M}
{Metzger}, B.~D. \& {Pejcha}, O. 2017, \mnras, 471, 3200

\bibitem[{{Meyer} \& {Meyer-Hofmeister}(1979)}]{mmh79}
{Meyer}, F. \& {Meyer-Hofmeister}, E. 1979, \aap, 78, 167

\bibitem[{{Miyaji} \& {Nomoto}(1987)}]{1987ApJ...318..307M}
{Miyaji}, S. \& {Nomoto}, K. 1987, \apj, 318, 307

\bibitem[{{Miyaji} {et~al.}(1980){Miyaji}, {Nomoto}, {Yokoi}, \&
  {Sugimoto}}]{1980PASJ...32..303M}
{Miyaji}, S., {Nomoto}, K., {Yokoi}, K., \& {Sugimoto}, D. 1980, \pasj, 32, 303

\bibitem[{{Moorwood} {et~al.}(1998){Moorwood}, {Cuby}, \&
  {Lidman}}]{moorwood1998}
{Moorwood}, A., {Cuby}, J.~G., \& {Lidman}, C. 1998, The Messenger, 91, 9

\bibitem[{{Morrell} {et~al.}(2014){Morrell}, {Contreras}, {Gonzalez},
  {Phillips}, {Hsiao}, {Stritzinger}, {Gall}, {Holmbo}, {Marples}, {Bock},
  {Parker}, {Drescher}, {Pearl}, {Taddia}, {Sanchez}, {Lira}, {Holoien},
  {Stanek}, {Kochanek}, {Davis}, {Basu}, {Beacom}, {Shappee}, {Prieto},
  {Bersier}, {Brimacombe}, {Szczygiel}, {Pojmanski}, {Hadjiyska}, {Walker},
  {Rabinowitz}, {Baltay}, {Ellman}, {McKinnon}, {Feindt}, \&
  {Nugent}}]{2014ATel.6508....1M}
{Morrell}, N., {Contreras}, C., {Gonzalez}, C., {et~al.} 2014, The Astronomer's
  Telegram, 6508

\bibitem[{{Nomoto}(1984)}]{1984ApJ...277..791N}
{Nomoto}, K. 1984, \apj, 277, 791

\bibitem[{{Nyholm} {et~al.}(2019){Nyholm}, {Sollerman}, {Tartaglia}, {Taddia},
  {Fremling}, {Blagorodnova}, {Filippenko}, {Gal-Yam}, {Howell},
  {Karamehmetoglu}, {Kulkarni}, {Laher}, {Leloudas}, {Masci}, {Kasliwal},
  {Mor{\r{a}}}, {Moriya}, {Ofek}, {Papadogiannakis}, {Quimby}, {Rebbapragada},
  \& {Schulze}}]{nyholm19}
{Nyholm}, A., {Sollerman}, J., {Tartaglia}, L., {et~al.} 2019, arXiv e-prints,
  arXiv:1906.05812

\bibitem[{{Osterbrock} \& {Ferland}(2006)}]{2006agna.book.....O}
{Osterbrock}, D.~E. \& {Ferland}, G.~J. 2006, {Astrophysics of gaseous nebulae
  and active galactic nuclei} (University Science Books)

\bibitem[{{Pastorello} {et~al.}(2019{\natexlab{a}}){Pastorello}, {Chen}, {Cai},
  {Morales-Garoffolo}, {Cano}, {Mason}, {Barsukova}, {Benetti}, {Berton},
  {Bose}, {Bufano}, {Callis}, {Cannizzaro}, {Cartier}, {Chen}, {Dong},
  {Dyrbye}, {Elias-Rosa}, {Fl{\"o}rs}, {Fraser}, {Geier}, {Goranskij}, {Kann},
  {Kuncarayakti}, {Onori}, {Reguitti}, {Reynolds}, {Losada}, {Sagu{\'e}s
  Carracedo}, {Schweyer}, {Smartt}, {Tatarnikov}, {Valeev}, {Vogl}, {Wevers},
  {de Ugarte Postigo}, {Izzo}, {Inserra}, {Kankare}, {Maguire}, {Smith},
  {Stalder}, {Tartaglia}, {Th{\"o}ne}, {Valerin}, \&
  {Young}}]{2019A&A...625L...8P}
{Pastorello}, A., {Chen}, T.~W., {Cai}, Y.~Z., {et~al.} 2019{\natexlab{a}},
  \aap, 625, L8

\bibitem[{{Pastorello} \& {Fraser}(2019)}]{2019NatAs...3..676P}
{Pastorello}, A. \& {Fraser}, M. 2019, Nature Astronomy, 3, 676

\bibitem[{{Pastorello} {et~al.}(2019{\natexlab{b}}){Pastorello}, {Mason},
  {Taubenberger}, {Fraser}, {Cortini}, {Tomasella}, {Botticella}, {Elias-Rosa},
  {Kotak}, {Smartt}, {Benetti}, {Cappellaro}, {Turatto}, {Tartaglia},
  {Djorgovski}, {Drake}, {Berton}, {Briganti}, {Brimacombe}, {Bufano}, {Cai},
  {Chen}, {Christensen}, {Ciabattari}, {Congiu}, {Dimai}, {Inserra}, {Kankare},
  {Magill}, {Maguire}, {Martinelli}, {Morales-Garoffolo}, {Ochner}, {Pignata},
  {Reguitti}, {Sollerman}, {Spiro}, {Terreran}, \&
  {Wright}}]{2019A&A...630A..75P}
{Pastorello}, A., {Mason}, E., {Taubenberger}, S., {et~al.} 2019{\natexlab{b}},
  \aap, 630, A75

\bibitem[{{Persson} {et~al.}(2013){Persson}, {Murphy}, {Smee}, {Birk},
  {Monson}, {Uomoto}, {Koch}, {Shectman}, {Barkhouser}, {Orndorff}, {Hammond},
  {Harding}, {Scharfstein}, {Kelson}, {Marshall}, \&
  {McCarthy}}]{2013PASP..125..654P}
{Persson}, S.~E., {Murphy}, D.~C., {Smee}, S., {et~al.} 2013, \pasp, 125, 654

\bibitem[{{Pettini} \& {Pagel}(2004)}]{2004MNRAS.348L..59P}
{Pettini}, M. \& {Pagel}, B. E.~J. 2004, \mnras, 348, L59

\bibitem[{{Phillips} {et~al.}(2019){Phillips}, {Contreras}, {Hsiao}, {Morrell},
  {Burns}, {Stritzinger}, {Ashall}, {Freedman}, {Hoeflich}, {Persson}, {Piro},
  {Suntzeff}, {Uddin}, {Anais}, {Baron}, {Busta}, {Campillay}, {Castell{\'o}n},
  {Corco}, {Diamond}, {Gall}, {Gonzalez}, {Holmbo}, {Krisciunas}, {Roth},
  {Ser{\'o}n}, {Taddia}, {Torres}, {Anderson}, {Baltay}, {Folatelli},
  {Galbany}, {Goobar}, {Hadjiyska}, {Hamuy}, {Kasliwal}, {Lidman}, {Nugent},
  {Perlmutter}, {Rabinowitz}, {Ryder}, {Schmidt}, {Shappee}, \&
  {Walker}}]{phillips2019}
{Phillips}, M.~M., {Contreras}, C., {Hsiao}, E.~Y., {et~al.} 2019, \pasp, 131,
  014001

\bibitem[{{Phillips} {et~al.}(2013){Phillips}, {Simon}, {Morrell}, {Burns},
  {Cox}, {Foley}, {Karakas}, {Patat}, {Sternberg}, {Williams}, {Gal-Yam},
  {Hsiao}, {Leonard}, {Persson}, {Stritzinger}, {Thompson}, {Campillay},
  {Contreras}, {Folatelli}, {Freedman}, {Hamuy}, {Roth}, {Shields}, {Suntzeff},
  {Chomiuk}, {Ivans}, {Madore}, {Penprase}, {Perley}, {Pignata}, {Preston}, \&
  {Soderberg}}]{2013ApJ...779...38P}
{Phillips}, M.~M., {Simon}, J.~D., {Morrell}, N., {et~al.} 2013, \apj, 779, 38

\bibitem[{{Pignata} {et~al.}(2009){Pignata}, {Maza}, {Hamuy}, {Antezana}, \&
  {Gonzales}}]{2009RMxAC..35R.317P}
{Pignata}, G., {Maza}, J., {Hamuy}, M., {Antezana}, R., \& {Gonzales}, L. 2009,
  in Revista Mexicana de Astronomia y Astrofisica Conference Series, Vol.~35,
  Revista Mexicana de Astronomia y Astrofisica Conference Series, 317

\bibitem[{{Podsiadlowski}(2001)}]{pod01}
{Podsiadlowski}, P. 2001, in Astronomical Society of the Pacific Conference
  Series, Vol. 229, Evolution of Binary and Multiple Star Systems, ed.
  {P.~Podsiadlowski, S.~Rappaport, A.~R.~King, F.~D'Antona, \& L.~Burderi },
  239--+

\bibitem[{{Poelarends} {et~al.}(2008){Poelarends}, {Herwig}, {Langer}, \&
  {Heger}}]{2008ApJ...675..614P}
{Poelarends}, A.~J.~T., {Herwig}, F., {Langer}, N., \& {Heger}, A. 2008, \apj,
  675, 614

\bibitem[{{Prieto} {et~al.}(2008){Prieto}, {Kistler}, {Thompson}, {Y{\"u}ksel},
  {Kochanek}, {Stanek}, {Beacom}, {Martini}, {Pasquali}, \&
  {Bechtold}}]{2008ApJ...681L...9P}
{Prieto}, J.~L., {Kistler}, M.~D., {Thompson}, T.~A., {et~al.} 2008, \apjl,
  681, L9

\bibitem[{{Prieto} {et~al.}(2009){Prieto}, {Sellgren}, {Thompson}, \&
  {Kochanek}}]{prieto2009}
{Prieto}, J.~L., {Sellgren}, K., {Thompson}, T.~A., \& {Kochanek}, C.~S. 2009,
  \apj, 705, 1425

\bibitem[{{Reichart} {et~al.}(2005){Reichart}, {Nysewander}, {Moran},
  {Bartelme}, {Bayliss}, {Foster}, {Clemens}, {Price}, {Evans}, {Salmonson},
  {Trammell}, {Carney}, {Keohane}, \& {Gotwals}}]{2005NCimC..28..767R}
{Reichart}, D., {Nysewander}, M., {Moran}, J., {et~al.} 2005, Nuovo Cimento C
  Geophysics Space Physics C, 28, 767

\bibitem[{{Retter} \& {Marom}(2003)}]{retter2003}
{Retter}, A. \& {Marom}, A. 2003, \mnras, 345, L25

\bibitem[{{Schlafly} \& {Finkbeiner}(2011)}]{2011ApJ...737..103S}
{Schlafly}, E.~F. \& {Finkbeiner}, D.~P. 2011, \apj, 737, 103

\bibitem[{{Schlegel}(1990)}]{1990MNRAS.244..269S}
{Schlegel}, E.~M. 1990, \mnras, 244, 269

\bibitem[{{Smartt} {et~al.}(2015){Smartt}, {Valenti}, {Fraser}, {Inserra},
  {Young}, {Sullivan}, {Pastorello}, {Benetti}, {Gal-Yam}, {Knapic},
  {Molinaro}, {Smareglia}, {Smith}, {Taubenberger}, {Yaron}, {Anderson},
  {Ashall}, {Balland}, {Baltay}, {Barbarino}, {Bauer}, {Baumont}, {Bersier},
  {Blagorodnova}, {Bongard}, {Botticella}, {Bufano}, {Bulla}, {Cappellaro},
  {Campbell}, {Cellier-Holzem}, {Chen}, {Childress}, {Clocchiatti},
  {Contreras}, {Dall'Ora}, {Danziger}, {de Jaeger}, {De Cia}, {Della Valle},
  {Dennefeld}, {Elias-Rosa}, {Elman}, {Feindt}, {Fleury}, {Gall},
  {Gonzalez-Gaitan}, {Galbany}, {Morales Garoffolo}, {Greggio}, {Guillou},
  {Hachinger}, {Hadjiyska}, {Hage}, {Hillebrandt}, {Hodgkin}, {Hsiao}, {James},
  {Jerkstrand}, {Kangas}, {Kankare}, {Kotak}, {Kromer}, {Kuncarayakti},
  {Leloudas}, {Lundqvist}, {Lyman}, {Hook}, {Maguire}, {Manulis}, {Margheim},
  {Mattila}, {Maund}, {Mazzali}, {McCrum}, {McKinnon}, {Moreno-Raya},
  {Nicholl}, {Nugent}, {Pain}, {Pignata}, {Phillips}, {Polshaw}, {Pumo},
  {Rabinowitz}, {Reilly}, {Romero-Ca{\~n}izales}, {Scalzo}, {Schmidt},
  {Schulze}, {Sim}, {Sollerman}, {Taddia}, {Tartaglia}, {Terreran},
  {Tomasella}, {Turatto}, {Walker}, {Walton}, {Wyrzykowski}, {Yuan}, \&
  {Zampieri}}]{smartt2015}
{Smartt}, S.~J., {Valenti}, S., {Fraser}, M., {et~al.} 2015, \aap, 579, A40

\bibitem[{{Smith} {et~al.}(2016){Smith}, {Andrews}, {Van Dyk}, {Mauerhan},
  {Kasliwal}, {Bond}, {Filippenko}, {Clubb}, {Graham}, {Perley}, {Jencson},
  {Bally}, {Ubeda}, \& {Sabbi}}]{2016MNRAS.458..950S}
{Smith}, N., {Andrews}, J.~E., {Van Dyk}, S.~D., {et~al.} 2016, \mnras, 458,
  950

\bibitem[{{Smith} {et~al.}(2009){Smith}, {Ganeshalingam}, {Chornock},
  {Filippenko}, {Li}, {Silverman}, {Steele}, {Griffith}, {Joubert}, {Lee},
  {Lowe}, {Mobberley}, \& {Winslow}}]{2009ApJ...697L..49S}
{Smith}, N., {Ganeshalingam}, M., {Chornock}, R., {et~al.} 2009, \apjl, 697,
  L49

\bibitem[{Smith {et~al.}(2011)Smith, Li, Silverman, Ganeshalingam, \&
  Filippenko}]{smith11}
Smith, N., Li, W., Silverman, J.~M., Ganeshalingam, M., \& Filippenko, A.~V.
  2011, MNRAS, 415, 773

\bibitem[{{Soker} \& {Tylenda}(2003)}]{soker2003}
{Soker}, N. \& {Tylenda}, R. 2003, \apjl, 582, L105

\bibitem[{{Soker}(2020)}]{soker2020}
{Soker}, N.  2020, \apj, 893, 20

\bibitem[{{Stritzinger} {et~al.}(2012){Stritzinger}, {Taddia}, {Fransson},
  {Fox}, {Morrell}, {Phillips}, {Sollerman}, {Anderson}, {Boldt}, {Brown},
  {Campillay}, {Castellon}, {Contreras}, {Folatelli}, {Habergham}, {Hamuy},
  {Hjorth}, {James}, {Krzeminski}, {Mattila}, {Persson}, \&
  {Roth}}]{2012ApJ...756..173S}
{Stritzinger}, M., {Taddia}, F., {Fransson}, C., {et~al.} 2012, \apj, 756, 173

\bibitem[Stritzinger et al.(2020)]{Stritzinger2020}
Stritzinger, M., Taddia, F.,  Fraser, M., et al. 2020, A\&A, in press (Paper I)

\bibitem[{{Taddia} {et~al.}(2013){Taddia}, {Stritzinger}, {Sollerman},
  {Phillips}, {Anderson}, {Boldt}, {Campillay}, {Castell{\'o}n}, {Contreras},
  {Folatelli}, {Hamuy}, {Heinrich-Josties}, {Krzeminski}, {Morrell}, {Burns},
  {Freedman}, {Madore}, {Persson}, \& {Suntzeff}}]{taddia13IIn}
{Taddia}, F., {Stritzinger}, M.~D., {Sollerman}, J., {et~al.} 2013, \aap, 555,
  A10

\bibitem[{{Thompson} {et~al.}(2009){Thompson}, {Prieto}, {Stanek}, {Kistler},
  {Beacom}, \& {Kochanek}}]{thompson09}
{Thompson}, T.~A., {Prieto}, J.~L., {Stanek}, K.~Z., {et~al.} 2009, \apj, 705,
  1364

\bibitem[{{Tully} {et~al.}(2013){Tully}, {Courtois}, {Dolphin}, {Fisher},
  {H{\'e}raudeau}, {Jacobs}, {Karachentsev}, {Makarov}, {Makarova},
  {Mitronova}, {Rizzi}, {Shaya}, {Sorce}, \& {Wu}}]{tully2013}
{Tully}, R.~B., {Courtois}, H.~M., {Dolphin}, A.~E., {et~al.} 2013, \aj, 146,
  86

\bibitem[{{Tully} {et~al.}(2016){Tully}, {Courtois}, \& {Sorce}}]{tully2016}
{Tully}, R.~B., {Courtois}, H.~M., \& {Sorce}, J.~G. 2016, \aj, 152, 50

\bibitem[{{Tylenda}(2005)}]{tylenda05}
{Tylenda}, R. 2005, \aap, 436, 1009

\bibitem[{{Tylenda} {et~al.}(2011){Tylenda}, {Hajduk}, {Kami{\'n}ski},
  {Udalski}, {Soszy{\'n}ski}, {Szyma{\'n}ski}, {Kubiak}, {Pietrzy{\'n}ski},
  {Poleski}, {Wyrzykowski}, \& {Ulaczyk}}]{tylenda11}
{Tylenda}, R., {Hajduk}, M., {Kami{\'n}ski}, T., {et~al.} 2011, \aap, 528, A114

\bibitem[{{Tylenda} \& {Soker}(2006)}]{tylenda2006}
{Tylenda}, R. \& {Soker}, N. 2006, \aap, 451, 223

\bibitem[{{Van Dyk} {et~al.}(2000){Van Dyk}, {Peng}, {King}, {Filippenko},
  {Treffers}, {Li}, \& {Richmond}}]{2000PASP..112.1532V}
{Van Dyk}, S.~D., {Peng}, C.~Y., {King}, J.~Y., {et~al.} 2000, \pasp, 112, 1532

\bibitem[{{Williams} {et~al.}(2015){Williams}, {Darnley}, {Bode}, \&
  {Steele}}]{wdbs15}
{Williams}, S.~C., {Darnley}, M.~J., {Bode}, M.~F., \& {Steele}, I.~A. 2015,
  \apjl, 805, L18

\end{thebibliography}
